%% file: main.tex
\documentclass[%
 amsmath,amssymb,
 aps,
]{revtex4-2}

\usepackage{graphicx}
\usepackage{dcolumn}
\usepackage{bm}

\begin{document}
    \title{Multi-mode quantum correlation generated from an unbalanced SU(1,1) interferometer using ultra-short laser pulses as pump}
    \author{Xueshi Guo$^1$}
    \email{xueshiguo@tju.edu.cn}
    \author{Wen Zhao$^1$}   
    \author{Xiaoying Li$^1$}
    \email{xiaoyingli@tju.edu.cn}
    \author{Z. Y. Ou$^2$}
    \email{jeffou@cityu.edu.hk}
\affiliation{%
$^{1}$College of Precision Instrument and Opto-Electronics Engineering, Key Laboratory of
Opto-Electronics Information Technology, Ministry of Education, Tianjin University,
Tianjin 300072, P. R. China\\
}
\affiliation{
$^{2}$ Department of Physics, City University of Hong Kong, 83 Tat Chee Avenue, Kowloon, Hong Kong, P. R. China\\
} 
    \date{\today}
             
\begin{abstract}
Multi-mode entanglement is one of the critical resource in quantum information technology. 
Generating large scale multi-mode entanglement state by coherently combining time-delayed continuous variables Einstein-Podolsky-Rosen pairs with linear beam-splitters has been widely studied recently. 
%
%
%
Here we theoretically investigate the multi-mode quantum correlation property of the optical fields generated from an unbalanced SU(1,1) interferometer pumped ultra-short pulses, which generates multi-mode entangled state by using a non-degenerate parametric processes to coherently combine delayed Einstein-Podolsky-Rosen pairs in different frequency band.
The covariance matrix of the generated multi-mode state is derived analytically for arbitrary mode number $M$ within adjacent timing slot, which shows a given mode is maximally correlated to 5 other modes. 
Based on the derived covariance matrix, both photon number correlation and quadrature amplitude correlation of the generated state is analyzed.
We also extend our analyzing method to the scheme of generating entangled state by using linear beam splitter as a coherent combiner of delayed EPR pairs, and compare the states generated by the two coherently combining schemes.
Our result provides a comprehensive theoretical description on the quantum correlations generated from an unbalanced SU(1,1) interferometer within Gaussian system range, and will offer more perspectives to quantum information technology.
\end{abstract}


\maketitle


\section{\label{sec:Introduction}Introduction}

Creating quantum correlation and entanglement is one of the crucial problems in quantum technology. 
In the perspective of quantum optics, a thorough description of a quantum optical system involves two part:  Firstly, the fast oscillation in optical frequency is described by a quantum harmonic oscillator. 
Secondly, all the other character of the system is attributed to optical modes, which is a normalized solution of Maxwell equations. 
Multi-mode quantum correlated state has been studied and demonstrated under different mode bases resident in spatial, temporal or polarization degree of freedom \cite{RevModPhys2020Treps}.
In recent years, time-division multiplexing generation of ultra-large scale entanglement state (ULSES) has been demonstrated by coherently combining time-delayed continuous variables (CV) Einstein-Podolsky-Rosen (EPR) pairs with beam-splitters (BS) \cite{Yokoyama2013, Furusawa2019Science, Larsen2019Science}.
In this scheme, the maximum mode number of the generated entangled state are only confined by the coherent length of the pump light generating the EPR pairs.
A commercial laser can have a spectral line-width of kilo-Hertz order, allowing the scale of generated entangled state up to 10$^6$ modes.

For the scheme of generating ULSES via time-division multiplexing, the principal requirement is to create correlation between optical modes in different timing slot by using coherent combinations. Besides beam-splitters, the coherent combinations can also be realized by using 
nonlinear parametric amplifiers,  which has been widely used in forming nonlinear interferometers.
%
As it is first shown in \cite{Yurke1986PRA}, a nonlinear interferometer is characterized by the lie group of SU(1,1) and the photon difference of the output two port is invariant. This is different from the conventional linear interferometer, which is characterized by the lie group of SU(2) and the photon summation of its two output port is invariant.
Because of this fundamental difference, nonlinear interferometers are also refereed to as SU(1,1) interferometers, and have been proved to possess merit over conventional linear interferometers when applied to quantum technologies such as quantum metrology, sensing, and quantum state engineering etc. \cite{Ou2020APLPhotonic}.
Very recent studies show an unbalanced SU(1,1) interferometer can be used as the coherent combination device of delayed EPR pairs to generate ultra-large scale entanglement state via time-division multiplexing \cite{Huo2022PRXQuantum, ZhouYanfen2023PRL}, and ULSES is successfully generated by using an unbalanced SU(1,1) interferometer which is based on continuous wave laser pumped non-degenerate four-wave mixing processes in hot $^{85}$Rb cell \cite{ZhouYanfen2023PRL}.  
%
Compared with the scheme using BS to realize a coherent combination, using unbalanced SU(1,1) interferometer allows the two modes of the EPR pairs in different frequency band and will generate ULSES with different correlation structure.

Besides, recent experimental work also shows an ultra-short pulse pumped parametric process is a good candidate for ULSES entangled state generation in time-division multiplexing manner \cite{Zhao2023OL,Huo2022PRXQuantum}. 
In such scheme, apart from the temporal mode resident in each timing slot, an individual ultra-short pulsed pump can generate signal and idler optical fields with multiple temporal-spectral modes, which are experimentally controllable and can serve an extra degree of freedom in the entanglement state generation process \cite{HuoNan2020PRL,DuPeilin2023OE}.
Ref \cite{Zhao2023OL} reported a quantum intensity correlation generated from pulse pumped parametric process using fiber as the nonlinear medium characterized in fast detection regime where each pulse can be distinguished by the detection process. Its result proves the in-dependency between generated quantum state in different timing slot.
Ref \cite{Huo2022PRXQuantum} reported an unbalanced SU(1,1) interferometer consists of two cascaded fiber optical parametric amplifiers (OPAs) pumping with pico-seconds pulses and an optical delay line in between. A measurement-dependent erasure of distinguish-ability is demonstrated in slow detection regime, where the response time of the detection process is many times larger than the time interval of the pump pulses. 
Apparently, the setup of unbalanced SU(1,1) using pulsed pump \cite{Huo2022PRXQuantum} can be used to generate ULSES when the detection process is fast enough to discriminate each optical pulse. 

From theoretical point of view, many toolboxes have been applied to the ultra-large scale entanglement state generation.
For the ULSES generated by using BS to realize coherent combination, complex graphical calculus representation \cite{Menicucci2011PRA} of the ULSES quantum state is derived, and the entanglement criterion based on linear combination of the quadratures is used \cite{VF2003PRA} to verify its entanglement property.
For the ULSES generated with an unbalanced SU(1,1) interferometer, the entanglement property for wave package units consist of 4 spatial-temporal modes is proven by using positive partial transpose (PPT) criterion \cite{Adesso2007JPA}, from which one can deduce any bi-partite division of the whole ULS optical state are entanglement.
However, for a multi-mode quantum state, the assertion of entangling existence between arbitrary two-partite division only gives limited information.
This is obvious when considering an extreme example: Imagining two mode (labeled with $\textbf{a}$ and $\textbf{b}$) are well entangled with each other, a two-partite entangled system (each partite labeled with $\textbf{S1}$ and $\textbf{S2}$) with arbitrary mode number can be formed as long as $\textbf{a} \in \textbf{S1}$ and $\textbf{b} \in \textbf{S2}$ is satisfied. 
For a Gaussian system, covariance matrix (CM) will give full information on the correlations between different modes, but the CM of the ULSES generated from an unbalanced SU(1,1) interferometer has not been derived.

In this work, we theoretically study the quantum correlation property of the state generated from an ultra-short pulse pumped unbalanced SU(1,1) interferometer in time-division multiplexing regime.
%
%
Defining the temporal modes by the timing slots of pump pulses, we derive the analytical form of the covariance matrix for the generated multi-mode state with arbitrary mode number $M$ in adjacent temporal modes. 
Our result shows that, when the pump is a pulse train with infinity pulse numbers, a single mode in an arbitrarily given timing slot is correlated with 5 other modes within 3 adjacent timing slots, and are independent with the modes outside these timing slots.
With the derived CM, we further study the multi-mode intensity correlation property and the quadrature entanglement property of the state.
%
%
%
Moreover, we extend the theoretical analysis method to the cluster state generation scheme in which linear beam splitter functions as the coherent combiner \cite{Yokoyama2013}, and compare the correlation structure between the ULSES realized by two different kinds of conherent combination devices.

The rest of the paper is organized as follows. We first introduce the theoretical model of an unbalanced SU(1,1) interferometer as a state preparation device In Sec. \ref{sec:thoery}, where temporal mode is defined and the timing order of non-degenerate parametric interactions is discussed. In Sec. \ref{sec:SU11}, we derive the covariance matrix of the state generate from a SU(1,1) interferometer, and discuss the intensity correlation and the quadrature entanglement property of the state. In Sec. \ref{sec:BS}, we extend our approach to the 1-D cluster state generation scheme in Ref. \cite{Yokoyama2013}, and compare the correlation properties of the quantum state generated in two different coherent combination approaches. Finally, we conclude in Sec. \ref{sec:summary}.


\section{\label{sec:thoery}An unbalanced SU(1,1) interferometer as a state preparation device}

\begin{figure}[h]
    \centering
    \includegraphics[width=0.90\textwidth]{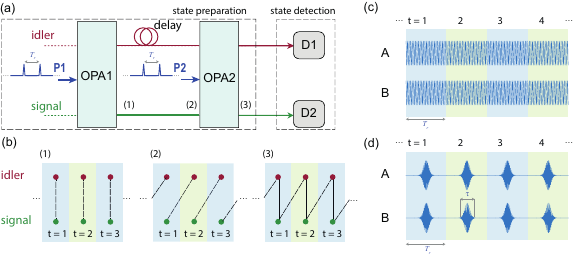}
    \caption{\label{schematic} 
    (a) The theoretical diagrams of an unbalanced SU(1,1) interferometer for multi-mode quantum correlated state generation, which consists of two cascaded OPAs each pumped by optical pulses (P1, P2) with uniform time interval $T_r$.
    The signal and idler injection are both vacuum states. 
    The delay line introduces a delay time the same as $T_r$.
    (b) The state preparation process of temporal multiplexed ultra large scale entanglement state. (c) The temporal mode definition for a continuous-wave pumped parametric process. (d) The temporal mode definition for a parametric process pumped by ultra-short pulses.
    }
\end{figure}

Our scheme for quantum state generation is shown in Fig.\ref{schematic}(a). It consists of two non-degenerate optical parametric amplifier (OPA1 and OPA2) pumped by ultra-short pulses (P1 and P2) with time interval $T_r$ and a delay line with a delay time exactly equals to $T_r$.
We assume the bandwidth of each frequency comb which forms the pump pulses is extremely narrow so the number of pump pluses within the coherence time of laser can approach to infinity. 
Without loss of generality, we put the delay line at the idler channel. Here we assume the detection process (D1 and D2) is fast enough to resolve each time slot defined by $T_r$.
As shown in Fig. \ref{schematic}(b), the state preparation process consists of 3 sub-stages:

(1). At the output of OPA1, multiple independent EPR pairs are generated with temporal mode defined by the pump pulses of OPA1 (P1). We label the temporal mode with an integer $t$. Lager index $t$ corresponds to the EPR pairs generated later in time;

(2). At the input of OPA2, delayed idler modes together with non-delayed signal modes are sent to the input of the OPA2 so they are coherently combined. For example, the idler mode generated by OPA1 at the timing slot $t=1$ is delayed, so it will be coherently combined with the signal mode at timing slot $t=2$ at the input of OPA2; 

(3). At the output of OPA2, the delayed modes are coherently combined by the two mode squeezing (TMSQ) operation in OPA2 so the ULSES is generated and sent to the state detection device. 

\subsection{Temporal mode definition}

We first make a comparison between the temporal mode definition for a continuous-wave (CW) pumped scheme and that for a pulse pumped scheme.
Fig.\ref{schematic}(c) shows the temporal mode definition of a CW pumped parametric process, where the blue curve represents the electrical field of the pump.
Since the pump is continuous, in the time-division multiplexing process one can arbitrarily define a timing slot length $T_r$, whose minimize value is only confined by the temporal resolution of the detection process.
As a comparison, Fig.\ref{schematic}(d) shows the temporal mode definition of an ultra-short pulse pumped parametric process.
In this case, the timing slot $T_r$ is the interval of the pump pulses, which is ultimately confined by the pulse duration of the pump pulses $\tau$.

Besides the temporal mode defined by $T_r$, a single ultra-shot pulse pumped parametric process has inherently multiple temporal-frequency mode property, which is closely related to the pulse duration $\tau$ and has been theoretically well studied \cite{RevModPhys2020Treps, Xueshi2015OE}. 
Following the procedure in Ref. \cite{Xueshi2015OE}, a non-degenerated parametric process pumped with a single pulse will generate a pair of continuous variable quantum entangled signal and idler pulses, whose spectral-temporal property can be modeled with an unitary operator $U_{s2}$ of a TMSQ operation with multiple temporal-spectral modes, which has the form of 
\begin{equation}
    \label{U_s2}
    \hat U_{s2} = \exp {[ r' e^{i\theta'} \iint F(\omega_s,\omega_i) \hat a^{\dagger}(\omega_s)\hat a^{\dagger}(\omega_i)d\omega_s d\omega_i -h.c. ]},
\end{equation}
where $\hat a^{\dagger}(\omega_s)$ and $\hat a^{\dagger}(\omega_i)$ as the creation operator of signal(idler) at the angular frequency $\omega_s$($\omega_i$), satisfying the commutation relationship $[\hat a_{s(i)}(\omega), \hat a_{s(i)}(\omega')] = \delta(\omega-\omega')$. 
$r'$ is a real number denoting the squeezing parameter, which is related to the intensity of the pump and the non-linearity of the gain media. $\theta'$ is the two-mode squeezing angle decide by the relative phase between pump and signal(idler) field.
$F(\omega_s,\omega_i)$ is a complex number valued joint spectral function of the parametric process, which is decided by the spectral property of pump and phase-matching condition of the parametric process. As it is noted in Ref. \cite{RevModPhys2020Treps}, it is always possible to define a group of independent EPR-like temporal-spectral mode pairs by using Bloch-Messiah-Williamson reduction, and $F(\omega_s,\omega_i)$ can be expanded with these mode pairs 
\begin{equation}
    \label{JSF}
	F\left(\omega_{s}, \omega_{i}\right)=\sum_{k} \xi_{k} e^{i\theta_{{k}}} \phi_{k}\left(\omega_{s}\right) \psi_{k}\left(\omega_{i}\right) \quad(k=1,2, \ldots).
\end{equation}
Using Eq.(\ref{JSF}), a group of independent temporal-spectral mode pair $\hat A_{ks}$ and $\hat A_{ki}$ can be defined as
\begin{eqnarray}
    \hat{A}_{(s,k)} & \equiv \int_{S} \phi_{k}^{*}\left(\omega_{s}^{\prime}\right) \hat{a}_{s}\left(\omega_{s}^{\prime}\right) d \omega_{s}^{\prime}  \nonumber\\
    \hat{A}_{(i,k)} & \equiv \int_{I} \psi_{k}^{*}\left(\omega_{i}^{\prime}\right) \hat{a}_{i}\left(\omega_{i}^{\prime}\right) d \omega_{i}^{\prime},
\end{eqnarray}
with commutation relation of standard bosons $[\hat{A}_{(s/i, k)}, \hat{A}^\dagger_{ (s/i, k)}] = 1$. The pulse pumped parametric process will transform the operators according to the mapping 
\begin{eqnarray}
    \label{STM}
    \hat A_{(s,k)} &\mapsto \hat U_{S2} \hat A_{(s,k)} \hat U^\dagger_{S2}=\cosh \left(r' \times \xi_{k}\right) \hat{A}_{(s,k)}+ e^{i(\theta'+\theta_{k})} \sinh \left( r' \times \xi_{k} \right) \hat{A}_{(i,k)}^{\dagger} \nonumber\\
    \hat A_{(s,k)} &\mapsto \hat U_{S2} \hat A_{(i,k)} \hat U^\dagger_{S2}=\cosh \left(r' \times \xi_{k}\right) \hat{A}_{(i,k)}+ e^{i(\theta'+\theta_{k})} \sinh \left( r'  \times \xi_{k} \right) \hat{A}_{(s,k)}^{\dagger},
\end{eqnarray}

For parametric process pumped by a train of optical pulses, it is straightforward to include timing slot mode index $t$ into the theoretical model as an extra degree of freedom, and the operator evolution in this case can be described by the mapping 
\begin{eqnarray}
    \label{STM_and_PITM_1}
    \hat{A}_{(s,k,t)} &\mapsto \cosh \left(r_{k, t} \right) \hat{A}_{(s,k,t)}+ e^{i\theta_{k,t}} \sinh \left( r_{k, t} \right) \hat{A}_{(i,k,t)}^{\dagger} \nonumber\\
    \label{STM_and_PITM_2}
    \hat{A}_{(i,k,t)} &\mapsto \cosh \left(r_{k, t}\right) \hat{A}_{(i,k,t)}+ e^{i\theta_{k,t}} \sinh \left(r_{k, t} \right) \hat{A}_{(s,k,t)}^{\dagger},
\end{eqnarray}
where $r_{k,t} = r'_t \times \xi_{k} $ and $\theta_{k,t} = \theta_t +\theta_{k} $ are the TMSQ amplitude and angle for temporal-spectral modes index $k$ and timing slot modes index $t$. 

Eq. (\ref{STM_and_PITM_2}) give the full description of the time mode for an OPA pumped by an ultra-short pulse train. 
However, to focus on the mode interaction between different timing slots, we introduce the following assumptions in our further analysis: 
(1) Since timing slot modes and temporal-spectral modes are independent degree of freedom in the temporal mode definition, we assume joint spectral function is  factorable so that the OPA has single temporal-spectral mode property \cite{Xueshi2015OE} and the temporal-spectral mode index $k$ can be omitted hereafter.
(2) We ignore the optical and the detection loss.
(3) We only consider the spontaneous case so that both the signal and the idler channel are vacuum input. We note a displacement in any mode will not change the quantum correlation property and can be well addressed by classical theory.
(4) We assume the pump pulses for each OPA are uniform, though we note it is possible to vary the TMSQ parameter for each timing slot by applying fast phase/amplitude to the pump pulses. 
With these simplification, we can use two parameters $r$ and $\theta$ to describe a parametric process, and the operator evolution due to parametric process for arbitrary timing slot index $t$ can be described by the following matrix form:
\begin{equation}
\label{eq:stable_OPA}
\left[ \begin{matrix} \hat{A}_{(s,t)}  & \hat{A}_{(i,t)} & \hat{A}^\dagger_{(s,t)} & \hat{A}^\dagger_{(i,t)} \end{matrix} \right]^\mathbf{T} 
\mapsto \mathbf{S}_{tmsq} \left[ \begin{matrix} \hat{A}_{(s,t)}  & \hat{A}_{(i,t)} & \hat{A}^\dagger_{(s,t)}  & \hat{A}^\dagger_{(i,t)}\end{matrix} \right] ^\mathbf{T}
\end{equation}
with
\begin{equation}
\label{eq:S_tmsq}
\mathbf{S}_{tmsq} = \left[\begin{matrix} 
                    \mu               &  0                 &  0               & e^{i\theta} \nu   \\
                       0              &  \mu               & e^{i\theta}\nu   & 0                 \\
                       0              &  e^{-i\theta} \nu  & \mu              & 0                 \\
                     e^{-i\theta}\nu  &  0                 & 0                & \mu   
                 \end{matrix}\right]
\end{equation}
and $\mu = \cosh r$ and $\nu = \sinh r$. As we will discuss in Eq. (18) in the next section, the matrix form for TMSQ operation in Eqs. (\ref{eq:stable_OPA}-\ref{eq:S_tmsq}) can be easily extended to quantum system with more than 2 modes.

\subsection{modes interactions of a cascaded parametric process with an optical delay line}

As it is shown in Fig.\ref{schematic}(b), a pump pulse of P1 at the timing slot $t_0$ will carry out TMSQ operation between vacuum modes indexed with $(s,t_0)$ and $(i, t_0)$ and a pump pulse of P2 at the timing slot $t$ will carry out TMSQ operation between the modes with the index $(s,t_0)$ and $(i, t_0+1)$. 
By using Eqs. (\ref{eq:stable_OPA}-\ref{eq:S_tmsq}) and considering the effect of the delay line, the unbalanced SU(1,1) interferometer as a state preparation device can be described by the following operator mapping:
\begin{eqnarray}
\label{eq:opa2out}
 \hat A_{(s,t)} \mapsto \mu_{1} \mu_{2} \hat A_{(s,t)} + \nu_{1} \nu_{2} e^{i \theta} \hat A_{ (s,t+1) } + \mu_{1} \nu_{2} e^{i \theta} \hat A_{(i, t+1)}^{\dagger}  + \nu_{1} \mu_{2}  \hat A_{(i, t)}^{\dagger}  \nonumber\\ 
 \hat A_{(i,t)} \mapsto \mu_{1} \mu_{2} \hat A_{(i, t+1)} + \nu_{1} \nu_{2} e^{i \theta} \hat A_{(i,t)} + \mu_{1} \nu_{2} e^{i \theta} \hat A_{(s,t)}^{\dagger} + \nu_{1} \mu_{2} \hat A_{(s, t+1)}^{\dagger},
\end{eqnarray}
where $\mu_{1(2)}=\cosh (r_{1(2) })$ and $\nu_{1(2)}=\sinh (r_{1(2)})$ are the amplitude gain of the OPA1/OPA2, and $\theta'= \theta+\theta_k$. Eq.(\ref{eq:opa2out}) clearly shows a cascaded parametric process with an optical delay line can create quantum correlations among optical pulses in different timing slot. 

\begin{figure}[h]
    \centering
    \includegraphics[width=0.80\textwidth]{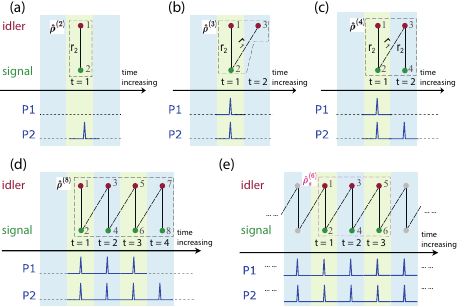}
    \caption{\label{pattern_illustration}
    \label{state_def}
    Modes interaction for generating quantum correlated optical state by using a one-pulse-delayed SU(1,1) interferometer. The squeezing parameter $r_1$ and $r_2$ for P1 and P2 are represented with dashed and solid line, respectively.
    (a) One pump pulse preforms TMSQ operation between one signal and one idler mode and generates quantum state with 2 modes; (b) Two pump pulses preform TMSQ between one signal and two idler modes and generate quantum state with 3 modes; (c) Three pump pulses preform TMSQ between two signal and two idler modes and generate quantum state with 4 modes; 
    In this way, $M-1$ pump pulses can generate quantum state with $M$ modes and we use $\hat \rho^{(M)}$ to represent the density operator of the state. The mode interaction of $\hat \rho^{(M)}$ for $M=8$ is depicted in (d). 
    When the number of pulses for both P1 and P2 trends to infinity, it is not possible to analyse the whole system. However, it is possible to analyse its subsystem having M modes in adjacent timing slot, and we use $\hat \rho_s^{(M)}$ to represent the density operator of this subsystem. (e) illustrates the mode interaction of $\hat \rho_s^{(M)}$ for the case of $M = 6$. As we will show later, $\hat \rho_s^{(M)}$ can be viewed as a subsystem of $\hat \rho^{(M+4)}$. 
    %
    }
\end{figure}

The scale and the structure of the generated state from an unbalanced SU(1,1) interferometer depends on the configuration of the pump pulses on both the number and the timing pattern. 
Specific to our state generation scheme, a state with $M$ optical modes resident in signal and idler channel need to be created by $M-1$ pump pulses.  
The three most simple cases for this rule is illustrated in Fig.\ref{state_def} (a) to (c). For the simplicity of the notation, we use an integer index to denote the different modes by the index mapping
\begin{eqnarray}
\label{mode_mapping}
    (i, t)   \mapsto & \ 2t-1 \ \  & \mathrm{for} \ \  t = 1, 2, 3 \cdots \nonumber\\
    (s, t)   \mapsto & \ 2t    & \mathrm{for} \ \ t = 1, 2, 3 \cdots.
\end{eqnarray}
The state shown in Fig.\ref{state_def} (a) is equivalent to a single pair of continuous variable Einstein-Podolsky-Rosen (EPR) pairs \cite{Zhao2023OL}, but with delay $\tau$ on the idler mode. The state shown in Fig.\ref{state_def} (b) and (c) are a time-division multiplexing realization of 3 or 4 modes entangled state equivalent to Ref. \cite{HailongWang2016OE, HailongWang2020OE}. Following the pattern of Fig.\ref{state_def} (a) to (c), we note the quantum system consists of $M$ optical mode by the operators $\hat \rho^{(M)}$ ($M\geq2$). For example, the state generated in Fig.\ref{state_def} (a) to (c) is noted with $\hat \rho^{(M)}$ for $M=2$, $M=3$ and $M=4$, respectively. 
To better illustrate the mode structure of $\rho^{(M)}$, we plot the mode interaction for $\hat \rho^{(8)}$ in Fig.\ref{pattern_illustration}(d), where the numbers of pump pulses for P1 and P2 are $3$ and $4$, respectively, creating correlations between 8 signal/idler modes within 4 timing slots. 

Since TMSQ operation to arbitrary two modes in a quantum system consists of $M$ modes is in general not commute, it is important to point out the timing order of the TMSQ operations. 
For the scheme in Fig.\ref{pattern_illustration}(d), the timing order of TMSQ operations follow the two rules: 

(1). Operations labelled by larger timing index $t$ always happen after that in a smaller timing index; 

(2). For the two operations connect to a given signal mode, which is represented with green dots and is indexed with and even number, the operation from OPA1 ($r_1$) always happens before that from OPA2 ($r_2$). 

\noindent By using the above rules, one can decide the timing order of TMSQ operations to derive $\rho^{(M)}$ for a given $M$. Taking $\hat \rho^{(8)}$ in Fig.\ref{state_def}(d) as an example, it can be viewed as a state generated by applying TMSQ operation to an 8-mode vacuum state in turns on mode pair (2,3) with OPA1, (1,2) with OPA2, (4,5) with OPA1, (3,4) with OPA2, (6,7) with OPA1, (5,6) with OPA2, and finally (7,8) with OPA2.
We note the idler field is delayed after the TMSQ operations from OPA1. Therefore, the dashed lines represent TMSQ oeprations of OPA1 is situated at the right side of its corresponding TMSQ operations from OPA2 (represented with solid lines), though it actually happens earlier. 

For time-division multiplexing scheme, one can easily increase the scale of the generated state by using more pump pulses within the coherent time of the laser used as pump, which can in principle hold millions of optical pulses. However, only a small amount of the modes can be analyzed due to the limitation of the state detection system. In this case, the number of the pump used can be regarded as infinity, and we can choose $M$ optical modes in $n$ successive timing slots to form a quantum system, whose density operator is noted by $\hat \rho_s^{(M)}$.  
Fig.\ref{pattern_illustration}(e) shows the mode interaction for $\hat \rho_s^{(6)}$ within 3 timing slot pumped by two stable optical pulse train. For a given mode number $M$, the major difference between $\hat \rho_s^{(M)}$ and $\hat \rho^{(M)}$ is that modes in $\hat \rho_s^{(M)}$ interacts with other modes (gray dots in Fig.\ref{pattern_illustration}(e)) outside the system but $\hat \rho^{(M)}$ does not interact with any mode outside the system.
As we will show later, the state  $\hat \rho_s^{(M)}$ can be viewed as a subsystem of $\hat \rho^{(M+4)}$, obtained by ignoring 4 modes each 2 from the earliest and the latest timing slot.

\section{\label{sec:SU11}Quantum Correlation properties derived in Phase space pictures}

\subsection{The derivation of covariance matrix}
Our scheme are fully within the Gaussian quantum state regime, so quantum correlation properties of both $\hat \rho^{(M)}$ and $\hat \rho_s^{(M)}$ can be derived in phase space picture, which has been systematically summarized in, for example, Ref. \cite{Adesso2014OpenSys}. A Gaussian state with mode number $M$ can be fully characterize by its displacement vector $\textbf{d}$ with $1\times2M$ in dimension
\begin{equation}
    \textbf{d} =  \left[ \langle{\hat A_1}\rangle,\langle{\hat A_2}\rangle ,  ... , \langle{\hat A_n}\rangle, \langle{\hat A^\dagger_1}\rangle, \langle{\hat A^\dagger_2}\rangle, ... , \langle{\hat A^\dagger_M}\rangle \right]^\textbf{T},
\end{equation}
and its covariance matrix (CM) in complex basis $\sigma_c$, which are $2M\times 2M$ in dimension and can be defined as
\begin{equation}
    \label{CM_def}
    \sigma_c = \left[\begin{matrix} \mathbf{A} & \mathbf{B} \\ \mathbf{ B}^* & \mathbf{ A}^* \end{matrix}\right],
\end{equation} 
with
\begin{eqnarray}
    A_{ij}  =& \langle \hat A_i \hat A^\dagger_j \rangle + \langle \hat A^\dagger_i \hat A_j  \rangle - 2 \langle{\hat A_i}\rangle \langle{\hat A_j^\dagger}\rangle  \nonumber\\
    B_{ij}  =& 2 \langle \hat A_i \hat A_j \rangle  - 2 \langle{\hat A_i}\rangle \langle{\hat A_j}\rangle
\end{eqnarray}
For an unbalanced SU(1,1) interferometer with vacuum input at both signal and idler mode, all the elements in the displacement vector $\textbf{d}$ satisfy $\langle{\hat A_i}\rangle = 0$ for $i=1$ to $2M$.
%
The evolution matrix of TMSQ in a $M$ mode system $\mathbf{S}_{tmsq}^{(M,i,j)}$ can be obtained by generalize the $\mathbf{S}_{tmsq}$ for 2-mode system in Eq.(\ref{eq:S_tmsq}) in the following step: 

(1). rewrite $\mathbf{S}_{tmsq}$ in Eq.(\ref{eq:S_tmsq}) into a $2 \times 2$ block matrix whose elements are all $2 \times 2$ sub-matrices; 

(2). Do direct sum to the diagonal sub-matrix with 1-dimension identity matrix, and direct sum the non-diagonal sub-matrix with 1-dimension zero matrix. Repeat both operation for $M-2$ times to get the new sub-matrices; 

(3). rearrange new sub-matrices according to the index $i$ and $j$.

\noindent Therefore, the evolution matrix of TMSQ for M mode system has a block matrix form of
\begin{equation}
    \label{S_tmsq}
    \mathbf{S}_{tmsq}^{(M,i,j)} = \left[\begin{matrix} \mathbf{S}_{A} & \mathbf{S}_{B} \\ \mathbf{ S}^*_{B} & \mathbf{ S}^*_{A} \end{matrix}\right],
\end{equation}
where $\mathbf{S}_{A}$ is diagonal matrix with the diagonal term of $S_{A_{ii}} = S_{A_{jj}} = \cosh(r)$ and $S_{A_{kk}} = 1$ for all $k\neq i$ or $k\neq j$, and $\mathbf{S}_{B}$ are $S_{B_{ij}} = S_{B_{ji}} = \sinh(r) e^{i\theta}$ and $S_{B_{kl}} = 0$ for all elements except $S_{B_{ij}}$ and $S_{B_{ji}}$.
%
The covariance matrices for $\hat \rho^{(M)}$ can be derived by apply the TMSQ operation to a M-mode vacuum state for $M-1$ rounds in the order we specified in Fig. \ref{state_def}, and the state evolution for each TMSQ can be represent by a $2M\times 2M$ matrix $\mathbf{S}_{tmsq}^{(M,i,j)}$
\begin{eqnarray}
    \label{state_evo}
   &\sigma_c   \mapsto \mathbf{S}_{tmsq}^{(M,i,j)} \sigma_c (\mathbf{S}_{tmsq}^{(M,i,j)})^\dagger 
\end{eqnarray}

We analytically calculate the matrix product in Eq. (\ref{state_evo}) with the help of the symbolic mathematics package Sympy \cite{sympy_citation}. 
By using Eqs. (\ref{S_tmsq}-\ref{state_evo}) multiple rounds, we can calculate covariance matrices $\sigma_c^{(M)}$ for the state $\hat \rho^{(M)}$. 
Starting from $M=2$, one can already clearly find the patterns for $\sigma_c^{(M)}$ of arbitrary mode number $M$ once the successively calculation is made up to $M\geq6$. 
Here we list the result for the state of $\rho^{(8)}$ in Fig.\ref{state_def}(d) and a general expression for arbitrary $M$ is given in Sec. I of the supplemental material. Using the notation of $\cosh(r_1) = \mu_1$, $\cosh(r_2) = \mu_2$, $\sinh(r_1) = \nu_1$, $\sinh(r_2) = \nu_1$, and $V_{1(2)} =\mu_{1(2)}^2 +\nu_{1(2)}^2 $, $c_{1(2)} = \mu_{1(2)}\nu_{1(2)}$, the $\mathbf{A}^{(8)}$ and $\mathbf{B}^{(8)}$ read
\begin{equation}
\label{CM_A8}
\mathbf{A}^{(8)} = \left[\begin{matrix}V_{1} \nu_{2}^{2} + \mu_{2}^{2} & 0 & 2 c_{1} c_{2} e^{i \theta} & 0 & 0 & 0 & 0 & 0\\0 & V_{1} \mu_{2}^{2} + \nu_{2}^{2} & 0 & 2 c_{1} c_{2} e^{- i \theta} & 0 & 0 & 0 & 0\\2 c_{1} c_{2} e^{- i \theta} & 0 & V_{1} V_{2} & 0 & 2 c_{1} c_{2} e^{i \theta} & 0 & 0 & 0\\0 & 2 c_{1} c_{2} e^{i \theta} & 0 & V_{1} V_{2} & 0 & 2 c_{1} c_{2} e^{- i \theta} & 0 & 0\\0 & 0 & 2 c_{1} c_{2} e^{- i \theta} & 0 & V_{1} V_{2} & 0 & 2 c_{1} c_{2} e^{i \theta} & 0\\0 & 0 & 0 & 2 c_{1} c_{2} e^{i \theta} & 0 & V_{1} V_{2} & 0 & 2 c_{1} c_{2} e^{- i \theta}\\0 & 0 & 0 & 0 & 2 c_{1} c_{2} e^{- i \theta} & 0 & V_{1} \mu_{2}^{2} + \nu_{2}^{2} & 0\\0 & 0 & 0 & 0 & 0 & 2 c_{1} c_{2} e^{i \theta} & 0 & V_{1} \nu_{2}^{2} + \mu_{2}^{2}\end{matrix}\right]
\end{equation}
and
\begin{equation}
\label{CM_B8}
\mathbf{B}^{(8)} = \left[\begin{matrix}0 & c_{2} \left(V_{1} + 1\right) e^{i \theta} & 0 & 2 c_{1} \nu_{2}^{2} e^{2 i \theta} & 0 & 0 & 0 & 0\\c_{2} \left(V_{1} + 1\right) e^{i \theta} & 0 & 2 c_{1} \mu_{2}^{2} & 0 & 0 & 0 & 0 & 0\\0 & 2 c_{1} \mu_{2}^{2} & 0 & 2 V_{1} c_{2} e^{i \theta} & 0 & 2 c_{1} \nu_{2}^{2} e^{2 i \theta} & 0 & 0\\2 c_{1} \nu_{2}^{2} e^{2 i \theta} & 0 & 2 V_{1} c_{2} e^{i \theta} & 0 & 2 c_{1} \mu_{2}^{2} & 0 & 0 & 0\\0 & 0 & 0 & 2 c_{1} \mu_{2}^{2} & 0 & 2 V_{1} c_{2} e^{i \theta} & 0 & 2 c_{1} \nu_{2}^{2} e^{2 i \theta}\\0 & 0 & 2 c_{1} \nu_{2}^{2} e^{2 i \theta} & 0 & 2 V_{1} c_{2} e^{i \theta} & 0 & 2 c_{1} \mu_{2}^{2} & 0\\0 & 0 & 0 & 0 & 0 & 2 c_{1} \mu_{2}^{2} & 0 & c_{2} \left(V_{1} + 1\right) e^{i \theta}\\0 & 0 & 0 & 0 & 2 c_{1} \nu_{2}^{2} e^{2 i \theta} & 0 & c_{2} \left(V_{1} + 1\right) e^{i \theta} & 0\end{matrix}\right]
\end{equation}

The results of $\sigma_c^{(M)}$ derived above can be used to further get the covariance matrices for the state $\hat \rho_s(M)$.
$M' \geq M+4$ is a sufficient condition to obtain the covariance matrices for the state $\hat \rho_s(M)$ by treating it as the subsystem of $\hat \rho(M')$.
In fact, Eqs. (\ref{CM_A8}-\ref{CM_B8}) show the first 2 and the last 2 boundary modes are special compared to the central modes, and all the other central modes are symmetric so that the CM does not change under mode index swapping. 
This rule also applies to all $\rho^{(M)}$ with an even number $M\geq6$. 
For $\rho^{(M)}$ with an odd mode number $M\geq5$, similar symmetric property exist but the special modes are the first 2 modes and the last 1 mode instead. 
To get the CM for $\hat \rho_s^{(M)}$, it is safe to trace out the first 2 and the last 2 modes of the CM for $\hat \rho^{(M+4)}$. The specific expressions of the covariance matrix for the state of $\rho_s^{(M)}$ is given in Sec. II of the supplemental material.  Therefore, the analytical result of CMs for both $\rho^{(M)}$ and $\rho_s^{(M)}$ for arbitrary mode number $M$ can be directly derived from our result, which serve as a full description of the quantum correlation property of the state and consist the main result of this work. 

Besides, Eqs.(\ref{CM_A8}-\ref{CM_B8}) indicate the state generation scheme in Fig.1 (a) can correlated the mode only 5 spatial-temporal adjacent modes within 3 timing slots. Mode pairs with time interval larger than 3 timing slot are independent since the corresponding correlation terms in both $\mathbf{A^{(8)}}$ and $\mathbf{B^{(8)}}$ are zero.
Therefore, the correlation property of $\rho_s^{(M)}$ for $M>6$ can be derived if we fully characterize the wave package of $M=6$.   

In the next, we use the result of CM to study the quantum correlation property of the optical state generated by the scheme in Fig.\ref{schematic}(a), including the intensity quantum correlation property of $\hat \rho^{(M)}$ and the quadrature entanglement property of $\hat \rho_s^{(M)}$.

\subsection{\label{Intensity} The Intensity Quantum Correlation Property} 
The non-degenerate parametric process always create photons in pairs, so the number of photon created in each of the mode in this process is always equal. 
This implies the state generate from an unbalanced SU(1,1) interferometer possess multi-mode intensity quantum correlation property. 
A general formalism for multi-mode photon number correlation property of a Gaussian system has been discussed in Ref. \cite{Vallone2019PRA}. 
The average photon number vector $\mathbf{m} = [ \langle \hat{N}_1 \rangle,  \langle \hat{N}_2 \rangle, \cdots, \langle \hat{N}_M \rangle ]^T$, with $ \hat N_i = \hat A_i^\dagger \hat A_i $, are the mean photon number of i$^{th}$ mode, are related to the CM in Eq. (\ref{CM_def}) with the equation of
\begin{equation}
\label{n_avg}
\langle \hat{N}_i \rangle =  \frac{1}{2}(A_{ii}-1) + |\langle \hat A_i \rangle|^2
\end{equation}
and the covariance matrix $\mathbf{K}$ for the photon number operator, whose element is defined as $K_{ij} \equiv  \langle \hat{N}_i \hat{N}_j \rangle - \langle \hat{N}_i \rangle \langle \hat{N}_j \rangle$, is related to Eq. (\ref{CM_def}) with the equation of
\begin{eqnarray}
\label{intensity_CM}
\mathbf{K} = &\frac{1}{4} ( \mathbf{A}\circ\mathbf{A}^* +  \mathbf{B}\circ\mathbf{B}^* - \mathbb{I}_M) + \nonumber\\ & \mathbf{Re}\Big[ (\mathbf{d} \mathbf{d}^T) \circ \mathbf{A} + (\mathbf{d} \mathbf{d}^\dagger) \circ \mathbf{B}\Big] 
\end{eqnarray} 
where $\circ$ denote the Hadamard product of matrices,
and $\mathbb{I}_M$ is a $M\times M$ identity matrix. We note the CM defined in Eq. (\ref{CM_def}) has a scaling factor of 2 compared to that in \cite{Vallone2019PRA}, therefore the $m_i$ in Eq.(\ref{n_avg}) and the $\mathbf{K}$ in Eq.(\ref{intensity_CM}) is scaled accordingly. 
%
Eq. (\ref{intensity_CM}) can be directly used to evaluate the variance of the linear combination of the photon number operator. By using the definition of covariance matrix $\mathbf{K}$, the variance of $\Delta( \sum_{k=1}^M \omega_k \hat{N}_k )^2$ can be written as a quadratic form of the parameters of the linear combination:
\begin{eqnarray}
\label{var_lin_N}
\Delta( \sum_{k=1}^M \omega_k \hat{N}_k )^2  = \bm{\omega} \mathbf{K} \bm{\omega}^T,
\end{eqnarray}
where $\bm{\omega} = [\omega_1, \omega_2, \cdots, \omega_M]^T$ is the parameter vector for the linear combination, and the square variance for a given operator $\hat O$ is defined as $\Delta\hat O ^2 = \langle \hat O ^2 \rangle - \langle\hat O\rangle ^2$.

By substituting the CMs we derive into Eqs. (\ref{n_avg}-\ref{var_lin_N}), the intensity correlation of both $\hat \rho^{(M)}$ and $\hat \rho_s^{(M)}$ can be characterized in terms of the mean photon number vector $\mathbf{m}$ and the covariance matrix for the photon number operator $\mathbf{K}$. 
We list these result in details in the supplemental material Sec. III. 
These calculation results show $\hat \rho_{(M)}$ having an ideal intensity correlation property that the variance of a particular linear combination of photon numbers in different mode are always vanishing. Specifically, for the state $\rho^{(M)}$, one can always find
\begin{equation}
\label{multi_mode_IDS_SU11}
\mathrm{for}\ \hat \rho^{(M)}: \ \ \ \  \Delta ( \sum_{k=1}^M (-1)^{k+1}\hat N_k)^2 = 0. 
\end{equation} 
This property is irrelevant of the parametric gain $r_1$, $r_2$ or the TMSQ phase $\theta$, and is originated from the fact that the unbalanced SU(1,1) interferometer has vacuum injections.
%
When the injection of the signal and/or idler is not vacuum but coherent states, the variance (\ref{multi_mode_IDS_SU11}) will be non-zero but still smaller than that of coherent states with the same average photon numbers, which is a multi-mode generalization of intensity difference squeezing reported in, for example, Ref. \cite{Zhao2023OL}. 
These modes can be separated into multiple places with high speed optical switches in principle. Therefore, the excellent quantum photon number correlation can be useful in different multi-user quantum information schemes.  

For the state $\hat \rho_s^{(M)}$, however, the photons generated in the parametric process can go into the boundary modes (the mode represented by the gray dots in Fig.\ref{pattern_illustration} (e)). Therefore, for the similar linear combination of photon numbers, the variance for the state $\hat \rho_s^{(M)}$ are
\begin{equation}
\label{multi_mode_IDS2}
\mathrm{for}\ \hat \rho_s^{(M)}: \ \ \ \  \Delta ( \sum_{k=1}^M (-1)^{k+1}\hat N_k)^2 = \Delta ( \hat N_2 - \hat N_1)^2 =  2\mu_1^2\nu_1^2. 
\end{equation} 
Eq. (\ref{multi_mode_IDS2}) shows, because of the difference in the boundary, the variance of the linear combination of photon number can not be zero. Therefore, to best utilize the photon correlation property of an unbalanced SU(1,1) interferometer, one need to chop the pump in order to prevent this boundary leakage of photon. 

\subsection{\label{Entanglement}The Quadrature Entanglement Property}

In this section, we investigate the quadrature entanglement property of the quantum states generated from an unbalanced SU(1,1) interferometer. 
We focus on the quadrature entanglement property for $\hat \rho_s^{(M)}$, where two different entanglement witnesses are used to quantitatively characterize the entanglement.  

Firstly, we study the variance of linear combinations of the quadrature operators in different modes. 
By using the basis changing matrices \cite{Adesso2014OpenSys}, the CM $\bm{\sigma}_c$ in complex basis can be changed into the quadrature operator basis $\bm{\sigma}$, whose elements are defined as 
\begin{equation}
\label{CM_xpxp}
\sigma_{i j}=\left\langle\hat{q}_i \hat{q}_j+\hat{q}_j \hat{q}_i\right\rangle-2\left\langle\hat{q}_i\right\rangle\left\langle\hat{q}_j\right\rangle,
\end{equation}
where $\hat{q}_i$ and $\hat q_j$ are the $i^{th}$ and $j^{th}$ elements in the quadrature vector $\bm{\widehat{q}}$ defined as
\begin{equation}
\bm{\hat{q}} = \Big[ \hat X_{1}, \hat P_{1}, \hat X_{2}, \hat P_{2}, \ldots , \hat X_{M}, \hat P_{M}  \Big]^T.
\end{equation}
Here for a given index $i$, the quadrature amplitude is defined as $ \hat X_{i} = (\hat A_i + \hat A^\dagger_i)/\sqrt{2}  $ and the quadrature phase is defined as $ \hat P_{i} = i( \hat A^\dagger_i - \hat A_i )/\sqrt{2} $. Therefore, the variance of a single vacuum mode is $ V_{sv} = 1/2$, and the shot noise limit for the linear combination of quadrature operators for $M$ vacuum modes are $ V_{snl}^{(M)} = M V_{sv}$. Using the CM in quadrature basis in Eq. (\ref{CM_xpxp}), the variance of linear combination for the quadratures $\Delta( \bm{\omega} \bm{\hat{q}} )^2$ are
\begin{eqnarray}
\label{var_lin_XP}
\Delta( \bm{\omega} \bm{\hat{q}} )^2 =  \frac{1}{2} \bm{\omega}^T \bm{\sigma} \bm{\omega}.
\end{eqnarray}
By using Eq. (\ref{var_lin_XP}), one can minimize the value of $\Delta( \bm{\omega} \bm{\hat{q}} )^2$ in shot noise unit by using different $\bm{\omega}$. For the 4 mode unity $\hat \rho_s^{(4)}$, we find the minimum noise for the linear combination are $V_{sv}=\frac{1}{4} V_{snl}^{(4)}$ in high gain limit. The specific value of $\bm{\omega}$ is related to the value of $\theta$ in the unbalanced SU(1,1) interferometer scheme, and for the special case of $\theta = 0$ and $r\to\infty$, the minimum variance of $\Delta( \bm{\omega} \bm{\hat{q}} )^2$ can be achieved with the following linear combination
\begin{eqnarray}
\label{LC_SU11_4}
\Delta(\hat X_{1} - \hat X_{2} + \hat X_{3} - \hat X_{4})^2 \to  V_{sv} = \frac{1}{4} V_{snl}^{(4)}  \\
\Delta(\hat P_{1} + \hat P_{2} + \hat P_{3} + \hat P_{4})^2 \to  V_{sv} = \frac{1}{4} V_{snl}^{(4)}.
\end{eqnarray}
This is different from the similar 4-mode wave-package unit generated with beam splitter in Ref. \cite{Yokoyama2013}, where the variance of the linear combination for 4 quadratures approaches to zero in high gain limit. We will give a more detailed comparison between the state generated from an unbalanced SU(1,1) interferometer and that generated from the scheme in Ref. \cite{Yokoyama2013} in the next subsection. Eq. (\ref{LC_SU11_4}) shows in 4-mode wave package unit, the squeezing rate of $\bm{\omega} \bm{\widehat{q}}$ operator is confined to less than 6 dB due to the difference coherent combination effect of a beam-splitter and a parametric process. However, this difference can be reduced when more modes goes into the wave-package. We minimum the variance of $\Delta( \bm{\omega} \bm{\hat{q}} )^2$ for $\hat \rho^{(M)}$ with even number of modes $M = 2m$ by varying the value of $\bm{\omega}$, and find the minimum variance of $\Delta( \bm{\omega} \bm{\hat{q}} )^2$ for $\theta = 0$ and $r\to\infty$ are
\begin{eqnarray}
\label{LC_SU11_M}
\Delta(\sum_{k=1}^M (-1)^{k+1}\hat X_k)^2 \to  V_{sv} = \frac{1}{M} V_{snl}^{(M)} \\
\Delta(\sum_{k=1}^M \hat P_k)^2 \to  V_{sv} = \frac{1}{M} V_{snl}^{(M)}
\end{eqnarray}
As Eqs. (27-28) shows, the minimum variance of the linear combination does not increase when modes are measured, and it keeps the value of vacuum noise of single mode. Therefore, when more modes is measured, the squeezing degree of $\Delta( \bm{\omega} \bm{\hat{q}} )^2$ can increase. This property is similar to the experimental result reported in Ref. \cite{Huo2022PRXQuantum} by using slow detector, which is equivalent to measuring more optical pulse pairs so that the visibility is recovered. 

We also investigate the bi-partite entanglement property for $\hat \rho_s^{(6)}$ by using PPT criterion knowing a given mode in $\hat \rho_s^{(M)}$ is only correlated to maximum 5 other modes. The number of different bi-partite subsystem pairs $n_6 $ can be found by using
\begin{equation}
n_6 = \sum_{k=1}^3 {{6}\choose{k}}{{6-k}\choose{k}}/2 + \sum_{k=1}^3\sum_{j=1+k}^{6-k} {{6}\choose{k}}{{6-k}\choose{j}} = 301. 
\end{equation}
We note here we allow the two subsystems (note with set A and B containing modes of $\hat \rho_s^{(6)}$ as their elements hereafter) do not have to cover all the 6 modes. If we add this constrain, however, the number of possible subsystem pairs are reduced to
\begin{equation}
n'_6 = {{6}\choose{1}} +  {{6}\choose{2}} +  {{6}\choose{3}}/2 = 31. 
\end{equation}
As it has been experimentally proven \cite{ZhouYanfen2023PRL}, all 31 possible subsystems pairs A and B exist some entanglement property when $A \cup B = \{ 1, 2, 3, 4, 5, 6 \} $. 
However, this doesn't apply to the $n_6-n'_6=270$ cases when $A \cup B \neq \{ 1, 2, 3, 4, 5, 6 \}$. We investigate this by following the procedure in Ref. \cite{Adesso2007JPA} and numerically calculate the minimum symplectic eign value of the partial transposed CM for different bi-partite divisions. These eign values are used to quantify the degree of entanglement between different bi-partite subsystem pairs A and B. To do this, we first rearrange the CM in quadrature operator basis $\bm{\sigma}$ into $\bm{\sigma}'_{A \mid B}$ according the divisions of the two partites, where the elements of $\bm{\sigma}'_{A \mid B}$ read
\begin{equation}
\sigma'_{i j}=\left\langle\hat{q'}_i \hat{q'}_j+\hat{q'}_j \hat{q'}_i\right\rangle-2\left\langle\hat{q'}_i\right\rangle\left\langle\hat{q'}_j\right\rangle
\end{equation}
with $\widehat{q}_i$ and $\widehat{q}_j$ are the i-th or j-th elements of the quadrature vector 
\begin{equation}
\bm{\hat{q'}} = \Big[ \underbrace{\hat X_{A_1}, \hat P_{A_1}, \hat X_{A_2}, \hat P_{A_2}, \ldots}_{2 N_A}  , \underbrace{\hat X_{B_1}, \hat P_{B_1}, \hat X_{B_2}, \hat P_{A_2}, \ldots}_{2 N_B}   \Big],
\end{equation}
where $X_{A(B)_j}$ and $P_{A(B)_j}$ are the quadrature amplitude and phase operators for the j-th mode in the partite A(B), and $N_A$($N_B$) is the number of modes in partite A(B). For the second step, we calculate the partial transposed CM $\bm{\tilde\sigma}_{A \mid B}$ according to the mode division
\begin{equation}
\bm{\tilde\sigma}_{A \mid B} \equiv \bm{\theta}_{A \mid B} \bm{\sigma'}_{A \mid B} \bm{\theta}_{A \mid B}
\end{equation}
with 
\begin{equation}
\boldsymbol{\theta}_{A \mid B}=\operatorname{diag}\{\underbrace{1,-1,1,-1, \ldots}_{2 N_A}, \underbrace{1,1,1,1, \ldots}_{2 N_B}\} .
\end{equation}
Finally, we numerically calculate 6 symplectic eign values $[\tilde \nu_1, \tilde \nu_2, \ldots, \tilde \nu_6]$ of $\bm{\tilde \sigma}_{A \mid B}$. According to PPT criterion, a necessary condition for partites $A$ and $B$ to be separable is $\tilde \nu_j \geq 1$ for $j=1$ to $6$ and we use the logarithmic minimum symplectic eign value 
\begin{equation}
L_{\mu}=\log_{10}(\min\{\tilde \nu_1, \tilde \nu_2, \ldots, \tilde \nu_6\})
\end{equation}
for entanglement quantification.

\begin{figure}[h]
    \centering
    \includegraphics[width=0.98\textwidth]{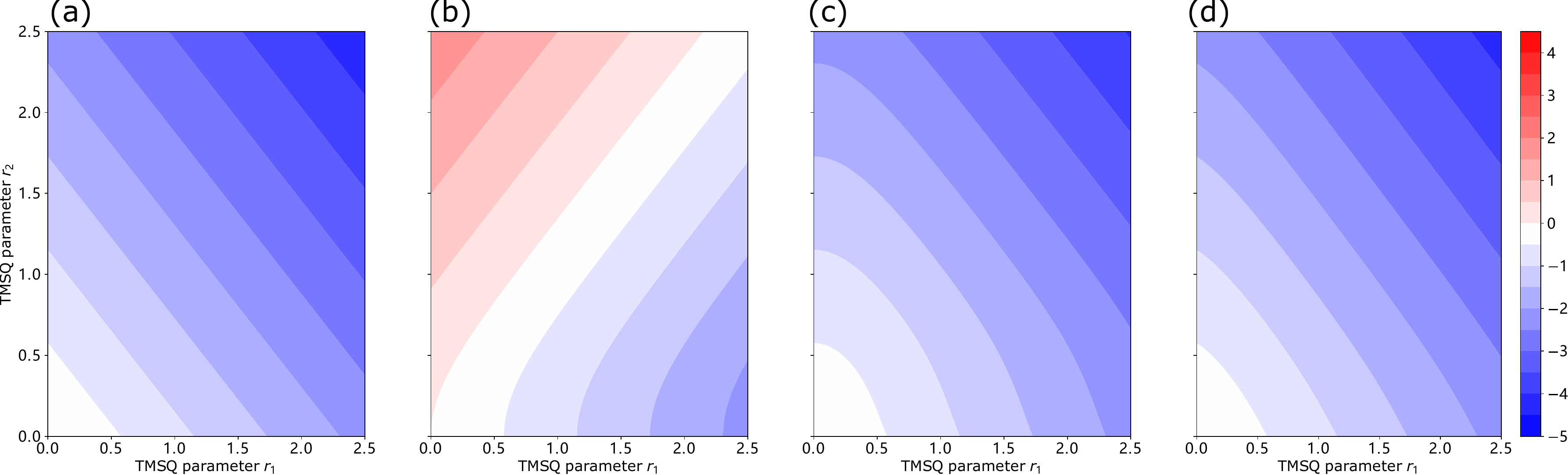}
    \caption{\label{6mode_PPT_1} 
    (a) The contour plot of minimum logarithmic PPT eign values for the CV EPR pairs generated from a balanced SU(1,1) in different squeezing parameter $r_1$ and $r_2$ for OPA1 and OPA2.
    %
    The contour plot of the minimum logarithmic PPT eign value for 3 special case of two-partite division consist of modes in $\hat \rho^{(6)}$ is plotted in (b), (c) and (d). The specific division of $A$ and $B$ are: (b), $A = \{1\}$ and $B = \{2\}$. (c), $A=\{1,3\}$ and $B = \{2,4\}$. (d), $A = \{1,3,5\}$ and ${B} = \{2,4,6\}$, respectively. 
    }
\end{figure}

Serving as a benchmark, we first calculate the $L_{\mu}$ for a pair of signal and idler pulses generated from a balanced SU(1,1) interferometer, which corresponds to the scheme in Fig.\ref{schematic}(a) when the delay is removed and both P1 and P2 are fully in phase ($\theta = 0$). The contour plot of $L_{\mu}$ for this case in different squeezing parameter for OPA1 and OPA2 $r_1$ and $r_2$ is shown in Fig.3 (a). One sees $L_{\mu}$ is always below zero indicating the entanglement property of signal and idler pulses, and the negativity of $L_{\mu}$ becomes more prominent as $r_1$ and $r_2$ increase. In the next, we calculate the $L_{\mu}$ values as a function of the TMSQ amplitude $r_1$ and $r_2$ when partite A consists of neighbouring signal pulses and partite B idler consists of corresponding neighbouring idler pulses, and the result is shown in Fig.3 (b)-(d). By comparing the result of Fig.3 (b)-(d) with that of Fig.3 (a), it is clear that when more signal/idler pulses is included in each partite, the entangle property of the quantum state generated from an unbalanced SU(1,1) asymptotically approaches to that generated from a cascade parametric process without delay. This result consists with the variance of the linear combination of quadratures in Eqs. (30-33). 

We further numerically calculate $L_{\mu}$ values as a function of the TMSQ amplitude $r_1$ and $r_2$ for all 301 different ways to divide 6 modes into two partites $A$ and $B$. As the result is long, we provide these results as a supplemental data and summarize the entanglement property in Sec. V of the Supplementary material. By summarizing these results, one sees the PPT negativity exist for all $r_1$ and $r_2$ values if the division of partites $A$ and $B$ satisfies any of the following condition: (1) A pair of nonempty subsets of $A$ and $B$ (noted as $A'$ and $B'$) can be found so that $A' \cup B'$ is two signal modes and two idler modes within two adjacent time slots; (2) $A$ and $B$ contains 2 and 3 modes, respectively. 
Apart from the two cases, the PPT negativity only exist partially for some $r_1$ and $r_2$ values, or even vanish for all $r_1$ and $r_2$ values. For example, Fig.3 (b) shows in the most range of $r_2>r_1$ the two partite $A = \{1\}$ and $B = \{2\}$ are not entangled with each other. More examples of partite division whose PPT negativity are partially exist or does not exist are given in Fig. 4. Particularly, we found two modes are always separable if no parametric amplification process is applied between them. 


\begin{figure}[h]
    \centering
    \includegraphics[width=0.98\textwidth]{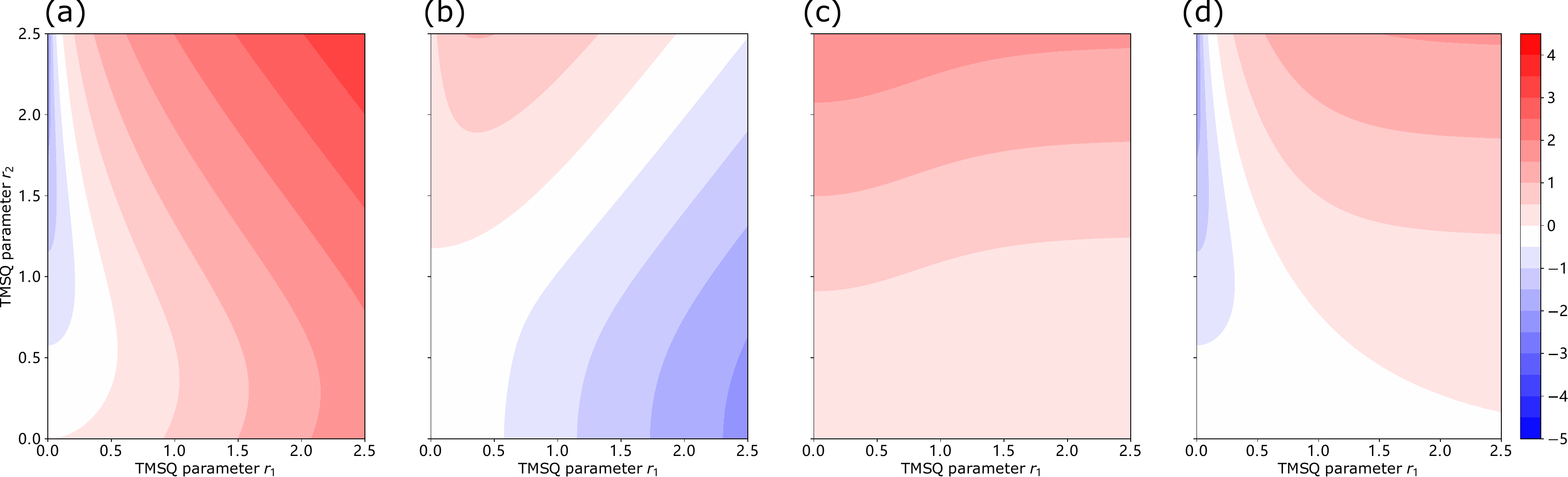}
    \caption{\label{6mode_PPT_2} The contour plot of the minimum logarithmic PPT eign value for 4 special case of two-partite division consist of modes in $\hat \rho^{(6)}$, which shows no negativity or only shows negativity in some special value of TMSQ parameter $r_1$ and $r_2$. The two parties $A$ and ${B}$ are: (a), $A = \{2\}$ and ${B} = \{3\}$. (b), $A = \{1\}$ and ${B} = \{2, 3\}$. (c), $A=\{1\}$ and ${B} = \{3,4\}$. (d), $A = \{2\}$ and ${B} = \{3,4\}$. 
    }
\end{figure}

\section{\label{sec:BS} Entangled state generated by combining delayed EPR pairs with a beam-splitter}

Here we compare the quantum correlation property of $\hat \rho^{(M)}$ and $\hat \rho_s^{(M)}$ generated from an unbalanced SU(1,1) nonlinear interferometer to the state with similarly parameter except for the OPA2 is replaced with a balanced beam splitter, which corresponds to the state generation scheme reported in \cite{Yokoyama2013}. The quantum correlation property of this case can be derived by using similar theoretical approach as we have presented in the above text. However, the TMSQ operation noted with $r_2$ in Fig.\ref{pattern_illustration} should be replaced with a 50:50 beam splitting operation with a phase parameter $\phi$, which can be modeled with the matrix of 
\begin{equation}
    \label{S_bs}
    \mathbf{S}_{bs}^{(M,i,j)} = \left[\begin{matrix} \mathbf{S}_{bs} & \mathbb{O}_M \\ \mathbb{O}_M & \mathbf{ S}^*_{bs} \end{matrix}\right],
\end{equation}
where the diagonal terms $\mathbf{S}_{bs}$ is  $S_{bs,{ii}} = \cos(\pi/4)$, 
$S_{bs,{jj}} = \cos(\pi/4) e^{i\phi} $  and $S_{T_{kk}} = 1$ for all $k\neq i$ or $k\neq j$. 
The non-zero non-diagonal terms are only $S_{bs,{ij}} = -\cos(\pi/4)$, $S_{bs,{ij}} = \cos(\pi/4) \mathrm{e}^{i\phi} $. $\mathbb{O}_M$ is a $M \times M$ zero matrix. Therefore, we can similarly define the quantum system with $M$ modes, which is noted by $\hat \rho_{bs}^{(M)}$, generated from $M-1$ operations of either TMSQ or beam splitting. With similar calculation, the CM for $\hat \rho_{bs}^{(M)}$ can be derived for a given $M$ in the form of Eq. (\ref{CM_def}), and the intensity correlation and the quadrature entanglement property can be analyzed similarly. These result for arbitrary M is given in Sec. IV. of the supplemental material and we only list the most import result to compare this case to the unbalanced SU(1,1) interferometer scheme.

Firstly, we list the case of $M=8$ to compare with the unbalanced SU(1,1) case in Eq. (\ref{CM_A8}-\ref{CM_B8}), which read
\begin{equation}
\label{A_bs_8}
\mathbf{A}^{(8)}_{bs} = \left[\begin{matrix}\frac{\mu_{1}^{2}}{2} + \frac{\nu_{1}^{2}}{2} + \frac{1}{2} & - \frac{\mu_{1}^{2}}{2} - \frac{\nu_{1}^{2}}{2} + \frac{1}{2} & 0 & 0 & 0 & 0 & 0 & 0\\- \frac{\mu_{1}^{2}}{2} - \frac{\nu_{1}^{2}}{2} + \frac{1}{2} & \frac{\mu_{1}^{2}}{2} + \frac{\nu_{1}^{2}}{2} + \frac{1}{2} & 0 & 0 & 0 & 0 & 0 & 0\\0 & 0 & V_{1} & 0 & 0 & 0 & 0 & 0\\0 & 0 & 0 & V_{1} & 0 & 0 & 0 & 0\\0 & 0 & 0 & 0 & V_{1} & 0 & 0 & 0\\0 & 0 & 0 & 0 & 0 & V_{1} & 0 & 0\\0 & 0 & 0 & 0 & 0 & 0 & \frac{\mu_{1}^{2}}{2} + \frac{\nu_{1}^{2}}{2} + \frac{1}{2} & \frac{\mu_{1}^{2}}{2} + \frac{\nu_{1}^{2}}{2} - \frac{1}{2}\\0 & 0 & 0 & 0 & 0 & 0 & \frac{\mu_{1}^{2}}{2} + \frac{\nu_{1}^{2}}{2} - \frac{1}{2} & \frac{\mu_{1}^{2}}{2} + \frac{\nu_{1}^{2}}{2} + \frac{1}{2}\end{matrix}\right]
\end{equation}
and
\begin{equation}
\label{B_bs_8}
\mathbf{B}^{(8)}_{bs} =\left[\begin{matrix}0 & 0 & - c_{1} e^{i \phi} & - c_{1} e^{i \phi} & 0 & 0 & 0 & 0\\0 & 0 & c_{1} e^{i \phi} & c_{1} e^{i \phi} & 0 & 0 & 0 & 0\\- c_{1} e^{i \phi} & c_{1} e^{i \phi} & 0 & 0 & - c_{1} e^{i \phi} & - c_{1} e^{i \phi} & 0 & 0\\- c_{1} e^{i \phi} & c_{1} e^{i \phi} & 0 & 0 & c_{1} e^{i \phi} & c_{1} e^{i \phi} & 0 & 0\\0 & 0 & - c_{1} e^{i \phi} & c_{1} e^{i \phi} & 0 & 0 & - c_{1} e^{i \phi} & - c_{1} e^{i \phi}\\0 & 0 & - c_{1} e^{i \phi} & c_{1} e^{i \phi} & 0 & 0 & c_{1} e^{i \phi} & c_{1} e^{i \phi}\\0 & 0 & 0 & 0 & - c_{1} e^{i \phi} & c_{1} e^{i \phi} & 0 & 0\\0 & 0 & 0 & 0 & - c_{1} e^{i \phi} & c_{1} e^{i \phi} & 0 & 0\end{matrix}\right],
\end{equation}
From Eq. (\ref{A_bs_8}), one firstly sees similar boundary effect exist and only the first and the last two modes are special. Therefore, we can similarly define the state $\hat \rho_{bs,s}^{(M)}$, which is generate by combining infinity pairs of delayed EPR state with a 50:50 beam splitter and taking out M adjacent modes out of it. We note $\hat \rho_{bs,s}^{(M)}$ 
is a good theoretical model for the 1-D cluster state generation scheme in Ref. \cite{Yoshikawa2016APLPhotonics}. Secondly, for a given non-boundary mode in $\hat \rho_{bs}^{(M)}$, there exists 4 rather than 5 other modes having non-zero correlation.  

Secondly, as it has been proved in Ref. \cite{Yoshikawa2016APLPhotonics}, for $\hat \rho_{bs,s}^{(4)}$, one can find the variance of the linear combinations of quadratures trends to zero when $r\to\infty$. For $\phi = 0$, these linear combination has the form of
\begin{eqnarray}
\Delta(\hat X_{1} + \hat X_{2} + \hat X_{3} -\hat X_{4})^2 \to 0\\
\Delta(\hat P_{1} + \hat P_{2} - \hat P_{3} +\hat P_{4})^2 \to 0.    
\end{eqnarray}
Compared to $\hat \rho_s^{(4)}$ generated from an unbalanced SU(1,1) interferometer, the variance of the linear combinations of quadratures for the $\hat \rho_{bs,s}^{(4)}$ case does not have the leakage noise with the amount of $V_{sv}$. We intuitively think the difference is due to more kind of operation can be used in the generation process of $\hat \rho_{bs,s}^{(M)}$. 

Further, in our formalism, the photon number quantum correlation property of $\hat \rho_{bs}^{(M)}$ can also be predicted. A TMSQ operation conserves the photon number difference between two modes being operated, but a beam splitting operation conserves photon number summation. This will make the correlation structure of the photon number for $\hat \rho^{(M)}$ and that for $\hat \rho_{bs}^{(M)}$ different. By substituting Eq. (\ref{A_bs_8}-\ref{B_bs_8}) into Eq. (\ref{intensity_CM}-\ref{var_lin_N}), one can find for $\hat \rho_{bs}^{(M)}$ with even mode number $M=2m$, the variance of linear combination of photon numbers in different mode are zero:
\begin{equation}
\label{multi_mode_IDS1}
\mathrm{for}\ \hat \rho_{bs}^{(M=2m)}: \ \ \ \  \Delta [ \sum_{k=1}^m (-1)^{k+1} (\hat N_{2k-1}+ \hat N_{2k})]^2 = 0. 
\end{equation} 
This relation is again originate from the photon number conservation, and is not related to the squeezing parameter $r_1$ or the phase $\phi$ introduced by the beam splitter. 

\section{\label{sec:summary} Summary}
In summary, we theoretically investigate the multi-mode quantum correlation structure of the optical fields generated from an unbalanced SU(1,1) nonlinear interferometer, which consist of two cascaded non-degenerate OPA and an optical delay line in between and can realize the coherent combination of two optical fields with different frequency. 
The covariance matrix of the generated state for the scheme is derived analytically for arbitrary mode number $M$ for two different pumping schemes: 
(1). The pump for the two OPA contains finite successive pulses and the signal and the idler fields generated by these pump pulses form a quantum system. In this case, $M-1$ pump pulses will generate a quantum state with $M$ optical modes, and the system is denoted by $\hat \rho^{(M)}$;
(2). The pump pulse trains for the two OPA contains infinite number of optical pulses, but only $M$ temporally adjacent signal/idler modes are studied as a subsystem, which is denoted by $\hat \rho_s^{(M)}$. 

\noindent The result shows that a given mode in such a state generation scheme is correlated to maximum 5 other modes and is independent to all the other modes. Besides, we show $\hat \rho_s^{(M)}$ with a mode number of $M$ can be viewed as a subsystem of $\hat \rho^{(M+4)}$. The intensity quantum correlation property and the quadrature entangle property of the generated optical modes are further analyzed and compared to the ultra-large scale entangled state generated by using a linear beam splitter to coherently combine the delayed EPR pairs.

Though we assume prefect detection efficiency in our theoretical analysis and the effect of optical loss is not include, it is straightforward to extend our theory to include the effect of the imperfection efficiency: An optical loss at a given mode $k$ of a $M$-mode quantum system can be modeled by adding an imaginary vacuum mode (with mode label $M+1$) and using the beam splitter operation in Eq. (\ref{S_bs}) between mode $k$ and mode $M+1$.
We finally note the unbalanced SU(1,1) interferometer here is working as a state preparation device, and this is different from using the second OPA of the SU(1,1) interferometer as part of the measurement device \cite{Jiamin2020PRA}. 
Therefore, the detection efficiency insensitive property for the parametric-amplifier-assisted homodyne detection scheme does not apply here.

Our result provides a comprehensive theoretical description on the quantum correlations generated from an unbalanced SU(1,1) interferometer within CV Gaussian system range, and will offer new perspectives to quantum information technology making use of multi-partite and multi-mode entanglement. The analyzing method used here may also be extended to other coherent combining approaches or multiplexing through other degree of freedom for optical fields.

\begin{acknowledgements}
	We would like to thank Prof. Qiongyi He for very instructive discussion. This work was supported in part by National Natural Science Foundation of China (Grants No.12004279).
\end{acknowledgements}

\bibliography{main}

\pagebreak
\widetext
\begin{center}
\textbf{\large Supplemental Materials for Distributed quantum sensing in a continuous variable entangled network}
\end{center}

\setcounter{equation}{0}
\setcounter{figure}{0}
\setcounter{table}{0}
\setcounter{page}{1}

\makeatletter
\renewcommand{\theequation}{S\arabic{equation}}
\renewcommand{\thefigure}{S\arabic{figure}}
\renewcommand{\bibnumfmt}[1]{[S#1]}
\renewcommand{\citenumfont}[1]{S#1}

\input{suppl_V2.tex}

\end{document}

%% file: suppl_V2.tex
\section{The covariance Matrix of the state $\hat \rho ^{(M)}$}
We use $\sigma_c^{(M)}$ to denote the covariance matrix (CM) of the Gaussian state $\rho^{(M)}$ (defined as Fig.2 (a)-(d) in the main text) generated from an unbalanced SU(1,1) interferometer pumping by $M-1$ pump pulses, and write $\sigma_c^{(M)}$ as a block matrix form:
\begin{equation}
	\label{CM_def}
	\sigma_c^{(M)} = \left[\begin{matrix} \mathbf{A}^{(M)} & \mathbf{B}^{(M)} \\ (\mathbf{ B}^{(M)})^* & (\mathbf{ A}^{(M)})^* \end{matrix}\right],
\end{equation}
According to the derivation of CM (see Section $\mathrm{\uppercase\expandafter{\romannumeral3}}$.A), the calculated $\sigma_c^{(M)}$ for $M=2$, $M=3$ and $M=4$ can be expressed in the following form:
\begin{align}
	\mathbf{A}^{(2)}= \left[\begin{matrix}V_{2} & 0\\0 & V_{2}\end{matrix}\right], 
	~~~
	\mathbf{B}^{(2)}=\left[\begin{matrix}0 & 2 c_{2} e^{i \theta}\\2 c_{2} e^{i \theta} & 0\end{matrix}\right],
\end{align}
\begin{align}
	\mathbf{A}^{(3)}= \left[\begin{matrix}V_{1} \nu_{2}^{2} + \mu_{2}^{2} & 0 & 2 c_{1} \nu_{2} e^{i \theta}\\0 & V_{1} \mu_{2}^{2} + \nu_{2}^{2} & 0\\2 c_{1} \nu_{2} e^{- i \theta} & 0 & V_{1}\end{matrix}\right],
	~~~
	\mathbf{B}^{(3)}=\left[\begin{matrix}0 & c_{2} \left(V_{1} + 1\right) e^{i \theta} & 0\\c_{2} \left(V_{1} + 1\right) e^{i \theta} & 0 & 2 c_{1} \mu_{2}\\0 & 2 c_{1} \mu_{2} & 0\end{matrix}\right],
\end{align}
and
\begin{align}
	\mathbf{A}^{(4)} = & \left[\begin{matrix}V_{1} \nu_{2}^{2} + \mu_{2}^{2} & 0 & 2 c_{1} c_{2} e^{i \theta} & 0\\0 & V_{1} \mu_{2}^{2} + \nu_{2}^{2} & 0 & 2 c_{1} c_{2} e^{- i \theta}\\2 c_{1} c_{2} e^{- i \theta} & 0 & V_{1} \mu_{2}^{2} + \nu_{2}^{2} & 0\\0 & 2 c_{1} c_{2} e^{i \theta} & 0 & V_{1} \nu_{2}^{2} + \mu_{2}^{2}\end{matrix}\right],
	\end{align}
\begin{align}
	\mathbf{B}^{(4)} = & \left[\begin{matrix}0 & c_{2} \left(V_{1} + 1\right) e^{i \theta} & 0 & 2 c_{1} \nu_{2}^{2} e^{2 i \theta}\\c_{2} \left(V_{1} + 1\right) e^{i \theta} & 0 & 2 c_{1} \mu_{2}^{2} & 0\\0 & 2 c_{1} \mu_{2}^{2} & 0 & c_{2} \left(V_{1} + 1\right) e^{i \theta}\\2 c_{1} \nu_{2}^{2} e^{2 i \theta} & 0 & c_{2} \left(V_{1} + 1\right) e^{i \theta} & 0\end{matrix}\right],
\end{align}
respectively. For $M=5$ and $M=6$,
\begin{align}
	\mathbf{A}^{(5)}=&\left[\begin{matrix}V_{1} \nu_{2}^{2} + \mu_{2}^{2} & 0 & 2 c_{1} c_{2} e^{i \theta} & 0 & 0\\0 & V_{1} \mu_{2}^{2} + \nu_{2}^{2} & 0 & 2 c_{1} c_{2} e^{- i \theta} & 0\\2 c_{1} c_{2} e^{- i \theta} & 0 & V_{1} V_{2} & 0 & 2 c_{1} \nu_{2} e^{i \theta}\\0 & 2 c_{1} c_{2} e^{i \theta} & 0 & V_{1} V_{2} & 0\\0 & 0 & 2 c_{1} \nu_{2} e^{- i \theta} & 0 & V_{1}\end{matrix}\right],
	\end{align}
\begin{align}
	\mathbf{B}^{(5)}=&\left[\begin{matrix}0 & c_{2} \left(V_{1} + 1\right) e^{i \theta} & 0 & 2 c_{1} \nu_{2}^{2} e^{2 i \theta} & 0\\c_{2} \left(V_{1} + 1\right) e^{i \theta} & 0 & 2 c_{1} \mu_{2}^{2} & 0 & 0\\0 & 2 c_{1} \mu_{2}^{2} & 0 & 2 V_{1} c_{2} e^{i \theta} & 0\\2 c_{1} \nu_{2}^{2} e^{2 i \theta} & 0 & 2 V_{1} c_{2} e^{i \theta} & 0 & 2 c_{1} \mu_{2}\\0 & 0 & 0 & 2 c_{1} \mu_{2} & 0\end{matrix}\right],
\end{align}
and
\begin{align}
	\mathbf{A}^{(6)}=&\left[\begin{matrix}V_{1} \nu_{2}^{2} + \mu_{2}^{2} & 0 & 2 c_{1} c_{2} e^{i \theta} & 0 & 0 & 0\\0 & V_{1} \mu_{2}^{2} + \nu_{2}^{2} & 0 & 2 c_{1} c_{2} e^{- i \theta} & 0 & 0\\2 c_{1} c_{2} e^{- i \theta} & 0 & V_{1} V_{2} & 0 & 2 c_{1} c_{2} e^{i \theta} & 0\\0 & 2 c_{1} c_{2} e^{i \theta} & 0 & V_{1} V_{2} & 0 & 2 c_{1} c_{2} e^{- i \theta}\\0 & 0 & 2 c_{1} c_{2} e^{- i \theta} & 0 & V_{1} \mu_{2}^{2} + \nu_{2}^{2} & 0\\0 & 0 & 0 & 2 c_{1} c_{2} e^{i \theta} & 0 & V_{1} \nu_{2}^{2} + \mu_{2}^{2}\end{matrix}\right],
	\end{align}
\begin{align}
	\mathbf{B}^{(6)}=&\left[\begin{matrix}0 & c_{2} \left(V_{1} + 1\right) e^{i \theta} & 0 & 2 c_{1} \nu_{2}^{2} e^{2 i \theta} & 0 & 0\\c_{2} \left(V_{1} + 1\right) e^{i \theta} & 0 & 2 c_{1} \mu_{2}^{2} & 0 & 0 & 0\\0 & 2 c_{1} \mu_{2}^{2} & 0 & 2 V_{1} c_{2} e^{i \theta} & 0 & 2 c_{1} \nu_{2}^{2} e^{2 i \theta}\\2 c_{1} \nu_{2}^{2} e^{2 i \theta} & 0 & 2 V_{1} c_{2} e^{i \theta} & 0 & 2 c_{1} \mu_{2}^{2} & 0\\0 & 0 & 0 & 2 c_{1} \mu_{2}^{2} & 0 & c_{2} \left(V_{1} + 1\right) e^{i \theta}\\0 & 0 & 2 c_{1} \nu_{2}^{2} e^{2 i \theta} & 0 & c_{2} \left(V_{1} + 1\right) e^{i \theta} & 0\end{matrix}\right].
\end{align}

With more calculation ,we find the analytical result of the CM $\sigma_c^{(M)}$ for $\rho^{(M)}$ ($M>4$) can be expressed in the following form: For odd number $M$, the diagonal elements in $\mathbf{A}^{(M)}$ has the form of
\begin{equation}
\label{eq:CM3}
\mathbf{A}^{(M)} (k, k)  = \left\{ 
\begin{array}{cl}
  V_1\nu_2^2+\mu_2^2 	&  \ \ \ \ k = 1 \\
  V_1\mu_2^2+\nu_2^2    &  \ \ \ \ k = 2 \\
  V_1V_2	            &  \ \ \ \ k = 3, 4, ..., M-1 \\
  V_1                   &  \ \ \ \ k = M   \\
\end{array}\right.
\end{equation}
and for even number $M$, the diagonal elements in $\mathbf{A}^{(M)}$ has the form of 
\begin{equation}
\mathbf{A}^{(M)} (k, k)  = \left\{ 
	\begin{array}{cl}
		V_1\nu_2^2+\mu_2^2 	&  \ \ \ \ k = 1 \  or \  k = M \\
		V_1\mu_2^2+\nu_2^2    &  \ \ \ \ k = 2 \  or \  k = M-1 \\
		V_1V_2	            &  \ \ \ \ k = 3, 4, ..., M-2 \\
	\end{array}\right.
\end{equation}
The non-zero off-diagonal elements of the $\mathbf{A}^{(M)}$ has the form of
\begin{equation}
\mathbf{A}^{(M)} (k+2, k)  = 2 c_1 c_2 \mathrm{e}^{(-1)^k i \theta} \ \ \ \ k = 1,2,...,M-2
\end{equation}
and 
\begin{equation}
\mathbf{A}^{(M)} (k, k+2)  = 2 c_1 c_2 \mathrm{e}^{(-1)^{k+1} i \theta} \ \ \ \ k = 1,2,...,M-2
\end{equation}
For odd number $M$, the non-zero off-diagonal elements of the matrix $\mathbf{B}^{(M)}$ can be expressed as
\begin{equation}
	\mathbf{B}^{(M)} (k+1, k) = \mathbf{B}^{(M)} (k, k+1)  = \left\{ 
	\begin{array}{cl}
		c_2(V_1+1)\mathrm{e}^{i\theta} 	&  \ \ \ \ k = 1 \\
		2\sin^2(k\pi/2) V_1c_2\mathrm{e}^{i\theta} + 2\cos^2(k\pi/2) c_1\mu_2^2  &  \ \ \ \ k = 2, 3, ..., M-2 \\
		2 c_1\mu_2 	&  \ \ \ \ k = M-1 
	\end{array}\right.
\end{equation}
and 
\begin{equation}
	\label{eq:CM8}
	\mathbf{B}^{(M)} (k+3, k) = \mathbf{B}^{(M)} (k, k+3)  = 2 \sin^2(k\pi/2) c_1 \nu_2^2 \mathrm{e}^{2i\theta} \ \ \ \ k = 1, 2, ... , M-3
\end{equation} 
For even number $M$, the non-zero off-diagonal elements of the matrix $\mathbf{B}^{(M)}$ can be expressed as
\begin{equation}
	\mathbf{B}^{(M)} (k+1, k) = \mathbf{B}^{(M)} (k, k+1)  = \left\{ 
	\begin{array}{cl}
		c_2(V_1+1)\mathrm{e}^{i\theta} 	&  \ \ \ \ k = 1 ~or ~ k = M-1 \\
		2\sin^2(k\pi/2) V_1c_2\mathrm{e}^{i\theta} + 2\cos^2(k\pi/2) c_1\mu_2^2  &  \ \ \ \ k = 2, 3, ..., M-2 
	\end{array}\right.
\end{equation}
and 
\begin{equation}
	\mathbf{B}^{(M)} (k+3, k) = \mathbf{B}^{(M)} (k, k+3)  = 2 \sin^2(k\pi/2) c_1 \nu_2^2 \mathrm{e}^{2i\theta} \ \ \ \ k = 1, 2, ... , M-3
\end{equation} 

\section{The covariance Matrix of the state $\hat \rho_s ^{(M)}$}

We use $\sigma_{c,s}^{(M)}$ to denote the covariance matrix (CM) of the Gaussian state $\rho_s^{(M)}$ (defined as Fig.2 (e) in the main text) generated from an unbalanced SU(1,1) interferometer pumping by $M-1$ pump pulses, and write $\sigma_{c,s}^{(M)}$ as a block matrix form:
\begin{equation}
	\label{CM_s_def}
	\sigma_{c,s}^{(M)} = \left[\begin{matrix} \mathbf{A_s}^{(M)} & \mathbf{B_s}^{(M)} \\ (\mathbf{ B_s}^{(M)})^* & (\mathbf{ A_s}^{(M)})^* \end{matrix}\right],
\end{equation}
The state $\rho^{(M)}_s$ can be viewed as a subsystem of $\rho^{(M+4)}$, by preserving the $M\times M$ sub-matrix in the middle of $\mathbf{A}^{(M+4)}$ and $\mathbf{B}^{(M+4)}$, we can get the block matrixs $\mathbf{A}^{(M)}_s$ and $\mathbf{B}^{(M)}_s$ of the CM $\sigma_{c,s}^{M}$ for $\rho^{(M)}_s$. The calculated CMs for $M=2$, $M=3$ and $M=4$ are given by
\begin{align}
   \mathbf{A}^{(2)}_s = \left[\begin{matrix}V_{1} V_{2} & 0\\0 & V_{1} V_{2}\end{matrix}\right],
   ~~~
   \mathbf{B}^{(2)}_s = \left[\begin{matrix}0 & 2 V_{1} c_{2} e^{i \theta}\\2 V_{1} c_{2} e^{i \theta} & 0\end{matrix}\right],
\end{align}
\begin{align}
	 \mathbf{A}^{(3)}_s = \left[\begin{matrix}V_{1} V_{2} & 0 & 2 c_{1} c_{2} e^{i \theta}\\0 & V_{1} V_{2} & 0\\2 c_{1} c_{2} e^{- i \theta} & 0 & V_{1} V_{2}\end{matrix}\right],
	 ~~~
	 \mathbf{B}^{(3)}_s = \left[\begin{matrix}0 & 2 V_{1} c_{2} e^{i \theta} & 0\\2 V_{1} c_{2} e^{i \theta} & 0 & 2 c_{1} \mu_{2}^{2}\\0 & 2 c_{1} \mu_{2}^{2} & 0\end{matrix}\right],
\end{align}
and
\begin{align}
	\mathbf{A}^{(4)}_s = \left[\begin{matrix}V_{1} V_{2} & 0 & 2 c_{1} c_{2} e^{i \theta} & 0\\0 & V_{1} V_{2} & 0 & 2 c_{1} c_{2} e^{- i \theta}\\2 c_{1} c_{2} e^{- i \theta} & 0 & V_{1} V_{2} & 0\\0 & 2 c_{1} c_{2} e^{i \theta} & 0 & V_{1} V_{2}\end{matrix}\right],
~~~
	\mathbf{B}^{(4)}_s = \left[\begin{matrix}0 & 2 V_{1} c_{2} e^{i \theta} & 0 & 2 c_{1} \nu_{2}^{2} e^{2 i \theta}\\2 V_{1} c_{2} e^{i \theta} & 0 & 2 c_{1} \mu_{2}^{2} & 0\\0 & 2 c_{1} \mu_{2}^{2} & 0 & 2 V_{1} c_{2} e^{i \theta}\\2 c_{1} \nu_{2}^{2} e^{2 i \theta} & 0 & 2 V_{1} c_{2} e^{i \theta} & 0\end{matrix}\right].
\end{align}
The non-zero elements of the block matrix $\mathbf{A}^{(M)}_s$ with mode number $M (M>3)$ can be expressed as
\begin{equation}
	\begin{aligned}
	\mathbf{A}^{(M)}_s(k,k) = & V_1V_2, \ \ \ \ k=1,2,...,M \\
	\mathbf{A}^{(M)}_s(k+2,k) = & 2 c_1 c_2 \mathrm{e}^{(-1)^k i \theta},  \ \ \ \ k = 1,2,...,M-2 \\
	\mathbf{A}^{(M)}_s(k,k+2) = & 2 c_1 c_2 \mathrm{e}^{(-1)^{k+1} i \theta}, \ \ \ \ k = 1,2,...,M-2 
	\end{aligned}
\end{equation}
The non-zero elements of the block matrix $\mathbf{B}^{(M)}_s$ with mode number $M (M>3)$ can be expressed as
\begin{equation}
		\mathbf{B}^{(M)}_s(k+1,k) =  \mathbf{B}^{(M)}_s(k,k+1) =
		2\sin^2(k\pi/2) V_1c_2\mathrm{e}^{i\theta} + 2\cos^2(k\pi/2) c_1\mu_2^2,  \ \ \ \ k = 1,2,...,M-1 \\ 
\end{equation}
\begin{equation}
	\mathbf{B}^{(M)}_s(k+3,k) = \mathbf{B}^{(M)}_s(k,k+3) =
	2 \sin^2(k\pi/2) c_1 \nu_2^2 \mathrm{e}^{2i\theta}, \ \ \ \ k = 1,2,...,M-3
\end{equation}

\section{The average photon number and the covariance matrix of photon number for the state of $\hat \rho ^{(M)}$ and $\hat \rho_s ^{(M)}$}

\subsection{ The result for the state of $\hat \rho ^{(M)}$}

The average photon number vector $\mathbf{m}^{(M)}$ of the state $\rho^{(M)}$ for $M=2$, $M=3$, and $M=4$ are
\begin{equation}
\mathbf{m}^{(2)} =  \left[\begin{matrix} \nu_{2}^{2} & \nu_{2}^{2}\end{matrix}\right]^T,
\end{equation}
\begin{equation}
\mathbf{m}^{(3)} = \left[\begin{matrix} \mu_{1}^{2} \nu_{2}^{2} & \mu_{1}^{2} \mu_{2}^{2} - 1 & \nu_{1}^{2}\end{matrix}\right]^T,
\end{equation}
and
\begin{equation}
\mathbf{m}^{(4)} = \left[\begin{matrix}\mu_{1}^{2} \nu_{2}^{2} & \mu_{1}^{2} \mu_{2}^{2} - 1 &  \mu_{1}^{2} \mu_{2}^{2} - 1 &  \mu_{1}^{2} \nu_{2}^{2}\end{matrix}\right]^T,
\end{equation}
respectively. For $M=5$ and $M=6$,
\begin{equation}
\mathbf{m}^{(5)} = \left[\begin{matrix}\mu_{1}^{2} \nu_{2}^{2} & \mu_{1}^{2} \mu_{2}^{2} - 1 & 0.5 V_{1} V_{2} - 0.5 & 0.5 V_{1} V_{2} - 0.5 & \nu_{1}^{2}\end{matrix}\right]^T,
\end{equation}
and
\begin{equation}
\mathbf{m}^{(6)} = \left[\begin{matrix}\mu_{1}^{2} \nu_{2}^{2} & \mu_{1}^{2} \mu_{2}^{2} - 1 & 0.5 V_{1} V_{2} - 0.5 & 0.5 V_{1} V_{2} - 0.5 & \mu_{1}^{2} \mu_{2}^{2} - 1 & \mu_{1}^{2} \nu_{2}^{2}\end{matrix}\right]^T.
\end{equation}
The analytical results of the average photon number vector $\mathbf{m}^{(M)}$ for $M>4$ can be written as
\begin{equation}
\mathbf{m}^{(M)} =  
\left[\begin{matrix}  \mathbf{m_h} & \mathbf{m_c} & \mathbf{m_t} \end{matrix}\right]^T
\end{equation}
with 
\begin{equation}
\mathbf{m_h} =  
\left[\begin{matrix}  \mu_{1}^{2} \nu_{2}^{2} & \mu_{1}^{2} \mu_{2}^{2} - 1  \end{matrix}\right],
\end{equation}
\begin{align}
\mathbf{m_c^{odd}} = \Big[ \underbrace{\frac{1}{2}(V_1V_2-1) ~~ \frac{1}{2}(V_1V_2-1) ~~ \cdots }_{ M - 3}  \Big],
 ~~~
\mathbf{m_c^{even}} = \Big[ \underbrace{\frac{1}{2}(V_1V_2-1) ~~ \frac{1}{2}(V_1V_2-1) ~~\cdots }_{ M - 4}  \Big],
\end{align}
and
\begin{align}
\mathbf{m_t^{odd}} =  
\left[\begin{matrix}  \nu_{1}^{2}  \end{matrix}\right],
 ~~~
\mathbf{m_t^{even}} =  
\left[\begin{matrix}  \mu_{1}^{2} \mu_{2}^{2} - 1  &  \mu_{1}^{2} \nu_{2}^{2} \end{matrix}\right],
\end{align}
where $\mathbf{m_c^{odd}}$ and $\mathbf{m_c^{even}}$ are the corresponding $\mathbf{m_c}$ with an odd number $M$ and an even number $M$, respectively. $\mathbf{m_t^{odd}}$ and $\mathbf{m_t^{even}}$ are the corresponding $\mathbf{m_t}$ with an odd number $M$ and an even number $M$, respectively.

For $M=2$, $M=3$, and $M=4$, the covariance matrix for the photon number operator $\mathbf{K}^{(M)}$ are
\begin{equation}
\mathbf{K}^{(2)} = \left[\begin{matrix}c_{2}^{2} & c_{2}^{2}\\c_{2}^{2} & c_{2}^{2}\end{matrix}\right],
\end{equation}
\begin{equation}
\mathbf{K}^{(3)} = \left[\begin{matrix}\mu_{1}^{4} \nu_{2}^{4} + \mu_{1}^{2} \nu_{2}^{2} & c_{2}^{2} \mu_{1}^{4} & c_{1}^{2} \nu_{2}^{2}\\c_{2}^{2} \mu_{1}^{4} & \mu_{1}^{4} \mu_{2}^{4} + \mu_{1}^{2} \mu_{2}^{2} & c_{1}^{2} \mu_{2}^{2}\\c_{1}^{2} \nu_{2}^{2} & c_{1}^{2} \mu_{2}^{2} & c_{1}^{2}\end{matrix}\right],
\end{equation}
and
\begin{equation}
\mathbf{K}^{(4)} = \left[\begin{matrix}\mu_{1}^{4} \nu_{2}^{4} + \mu_{1}^{2} \nu_{2}^{2} & c_{2}^{2} \mu_{1}^{4} & c_{1}^{2} c_{2}^{2} & c_{1}^{2} \nu_{2}^{4}\\c_{2}^{2} \mu_{1}^{4} & \mu_{1}^{4} \mu_{2}^{4} + \mu_{1}^{2} \mu_{2}^{2} & c_{1}^{2} \mu_{2}^{4} & c_{1}^{2} c_{2}^{2}\\c_{1}^{2} c_{2}^{2} & c_{1}^{2} \mu_{2}^{4} & \mu_{1}^{4} \mu_{2}^{4} + \mu_{1}^{2} \mu_{2}^{2} & c_{2}^{2} \mu_{1}^{4}\\c_{1}^{2} \nu_{2}^{4} & c_{1}^{2} c_{2}^{2} & c_{2}^{2} \mu_{1}^{4} & \mu_{1}^{4} \nu_{2}^{4} + \mu_{1}^{2} \nu_{2}^{2}\end{matrix}\right],
\end{equation}
respectively.
For $M=5$ and $M=6$, the covariance matrix for the photon number operator $\mathbf{K}^{(M)}$  are
\begin{equation}
\mathbf{K}^{(5)} = \left[\begin{matrix}\mu_{1}^{4} \nu_{2}^{4} + \mu_{1}^{2} \nu_{2}^{2} & c_{2}^{2} \mu_{1}^{4} & c_{1}^{2} c_{2}^{2} & c_{1}^{2} \nu_{2}^{4} & 0\\c_{2}^{2} \mu_{1}^{4} & \mu_{1}^{4} \mu_{2}^{4} + \mu_{1}^{2} \mu_{2}^{2} & c_{1}^{2} \mu_{2}^{4} & c_{1}^{2} c_{2}^{2} & 0\\c_{1}^{2} c_{2}^{2} & c_{1}^{2} \mu_{2}^{4} & 0.25 V_{1}^{2} V_{2}^{2} - 0.25 & V_{1}^{2} c_{2}^{2} & c_{1}^{2} \nu_{2}^{2}\\c_{1}^{2} \nu_{2}^{4} & c_{1}^{2} c_{2}^{2} & V_{1}^{2} c_{2}^{2} & 0.25 V_{1}^{2} V_{2}^{2} - 0.25 & c_{1}^{2} \mu_{2}^{2}\\0 & 0 & c_{1}^{2} \nu_{2}^{2} & c_{1}^{2} \mu_{2}^{2} & c_{1}^{2}\end{matrix}\right],
\end{equation}
and 
\begin{equation}
\mathbf{K}^{(6)} = \left[\begin{matrix}\mu_{1}^{4} \nu_{2}^{4} + \mu_{1}^{2} \nu_{2}^{2} & c_{2}^{2} \mu_{1}^{4} & c_{1}^{2} c_{2}^{2} & c_{1}^{2} \nu_{2}^{4} & 0 & 0\\c_{2}^{2} \mu_{1}^{4} & \mu_{1}^{4} \mu_{2}^{4} + \mu_{1}^{2} \mu_{2}^{2} & c_{1}^{2} \mu_{2}^{4} & c_{1}^{2} c_{2}^{2} & 0 & 0\\c_{1}^{2} c_{2}^{2} & c_{1}^{2} \mu_{2}^{4} & 0.25 V_{1}^{2} V_{2}^{2} - 0.25 & V_{1}^{2} c_{2}^{2} & c_{1}^{2} c_{2}^{2} & c_{1}^{2} \nu_{2}^{4}\\c_{1}^{2} \nu_{2}^{4} & c_{1}^{2} c_{2}^{2} & V_{1}^{2} c_{2}^{2} & 0.25 V_{1}^{2} V_{2}^{2} - 0.25 & c_{1}^{2} \mu_{2}^{4} & c_{1}^{2} c_{2}^{2}\\0 & 0 & c_{1}^{2} c_{2}^{2} & c_{1}^{2} \mu_{2}^{4} & \mu_{1}^{4} \mu_{2}^{4} + \mu_{1}^{2} \mu_{2}^{2} & c_{2}^{2} \mu_{1}^{4}\\0 & 0 & c_{1}^{2} \nu_{2}^{4} & c_{1}^{2} c_{2}^{2} & c_{2}^{2} \mu_{1}^{4} & \mu_{1}^{4} \nu_{2}^{4} + \mu_{1}^{2} \nu_{2}^{2}\end{matrix}\right].
\end{equation}

For an odd number $M ~(M>4)$, the non-zero elements of the CM $\mathbf{K}^{(M)}$ are
\begin{equation}
	\mathbf{K}^{(M)} (k, k)  = \left\{ 
	\begin{array}{cl}
		\mu_1^4\nu_2^4+\mu_1^2\nu_2^2 	&  \ \ \ \ k = 1 \\
		\mu_1^4\mu_2^4+\mu_1^2\mu_2^2    &  \ \ \ \ k = 2 \\
		\frac{1}{4}(V_1^2V_2^2-1)	    &  \ \ \ \ k = 3, 4, ..., M-1 \\
	  	c_1^2           &  \ \ \ \ k = M 
	\end{array}\right.
\end{equation}
\begin{equation}
	\mathbf{K}^{(M)} (k+1, k) = \mathbf{K}^{(M)} (k, k+1)  = \left\{ 
	\begin{array}{cl}
		c_2^2\mu_1^4 	&  \ \ \ \ k = 1 \\
		\sin^2(k\pi/2) V_1^2c_2^2 + \cos^2(k\pi/2) c_1^2\mu_2^4  &  \ \ \ \ k = 2, 3, ..., M-1 
	\end{array}\right.
\end{equation}
\begin{equation}
	\mathbf{K}^{(M)} (k+2, k) = \mathbf{K}^{(M)} (k, k+2)  = \left\{ 
	\begin{array}{cl}
		c_1^2c_2^2 	&  \ \ \ \ k = 1, 2, ..., M-3 \\
		c_1^2\nu_2^2  &  \ \ \ \ k = M-2 
	\end{array}\right.
\end{equation}
\begin{equation}
	\mathbf{K}^{(M)} (k+3, k) = \mathbf{K}^{(M)} (k, k+3)  = 
		\sin^2(k\pi/2)	c_1^2\nu_2^4    \ \ \ \ k = 1, 2, ..., M-3 
	\end{equation}

For an even number $M ~(M>4)$, the non-zero elements of the CM $\mathbf{K}^{(M)}$ are
\begin{equation}
	\mathbf{K}^{(M)} (k, k)  = \left\{ 
	\begin{array}{cl}
		\mu_1^4\nu_2^4+\mu_1^2\nu_2^2 	&  \ \ \ \ k = 1 ~or ~ k = M\\
		\mu_1^4\mu_2^4+\mu_1^2\mu_2^2    &  \ \ \ \ k = 2 ~or ~k = M-1\\
		\frac{1}{4}(V_1^2V_2^2-1)	    &  \ \ \ \ k = 3, 4, ..., M-2 
	\end{array}\right.
\end{equation}
\begin{equation}
	\mathbf{K}^{(M)} (k+1, k) = \mathbf{K}^{(M)} (k, k+1)  = \left\{ 
	\begin{array}{cl}
		c_2^2\mu_1^4 	&  \ \ \ \ k = 1 ~or ~k = M-1\\
		\sin^2(k\pi/2) V_1^2c_2^2 + \cos^2(k\pi/2) c_1^2\mu_2^4  &  \ \ \ \ k = 2, 3, ..., M-2 
	\end{array}\right.
\end{equation}
\begin{equation}
	\mathbf{K}^{(M)} (k+2, k) = \mathbf{K}^{(M)} (k, k+2)  = 
		c_1^2c_2^2 	  \ \ \ \ k = 1, 2, ..., M-2 
\end{equation}
\begin{equation}
	\mathbf{K}^{(M)} (k+3, k) = \mathbf{K}^{(M)} (k, k+3)  = 
		\sin^2(k\pi/2)	c_1^2\nu_2^4    \ \ \ \ k = 1, 2, ..., M-3 
	\end{equation}
 
\subsection{ The result for the state of $\hat \rho_s ^{(M)}$} 
 
For the state $\rho^{(M)}_s$, the mean photon number vector $\mathbf{m}^{(M)}_s$ with an arbitrary mode number $M ~(M\geq 2)$ has the form of
\begin{equation}
\mathbf{m}^{(M)}_s = \Big[ \underbrace{\frac{1}{2}(V_1V_2-1) ~~ \frac{1}{2}(V_1V_2-1) ~~\cdots }_{ M }  \Big].
\end{equation}

For the state $\rho^{(M)}_s$, the non-zero elements of the covariance matrix $\mathbf{K}^{(M)}_s$ for the photon number operator with $M=2$ and $M=3$ are
\begin{equation}
  \mathbf{K}^{(2)}_s =  \left[\begin{matrix}\frac{1}{4}(V_1^2V_2^2-1) & V_{1}^{2} c_{2}^{2}\\V_{1}^{2} c_{2}^{2} & \frac{1}{4}(V_1^2V_2^2-1)\end{matrix}\right],
\end{equation}
and
\begin{equation}
  \mathbf{K}^{(3)}_s =  \left[\begin{matrix}\frac{1}{4}(V_1^2V_2^2-1) & V_{1}^{2} c_{2}^{2} & c_{1}^{2} c_{2}^{2}\\V_{1}^{2} c_{2}^{2} & \frac{1}{4}(V_1^2V_2^2-1) & c_{1}^{2} \mu_{2}^{4}\\c_{1}^{2} c_{2}^{2} & c_{1}^{2} \mu_{2}^{4} & \frac{1}{4}(V_1^2V_2^2-1)\end{matrix}\right], 
\end{equation}
respectively.
For the state $\rho^{(M)}_s$, the non-zero elements of the covariance matrix $\mathbf{K}^{(M)}_s$ for the photon number operator with an arbitrary mode number $M >3$ can be given by
\begin{equation}
	\mathbf{K}^{(M)}_s (k, k)  =
	\frac{1}{4}(V_1^2V_2^2-1)	   \ \ \ \ k = 1, 2, ..., M 
	 \end{equation}
\begin{equation}
	\mathbf{K}^{(M)}_s (k+1, k) = \mathbf{K}^{(M)}_s (k, k+1)  = 	\sin^2(k\pi/2) V_1^2c_2^2 + \cos^2(k\pi/2) c_1^2\mu_2^4    \ \ \ \ k = 1, 2, ..., M-1 
\end{equation}
\begin{equation}
	\mathbf{K}^{(M)}_s (k+2, k) = \mathbf{K}^{(M)}_s (k, k+2)  = 	c_1^2c_2^2 	  \ \ \ \ k = 1, 2, ..., M-2 
\end{equation}
\begin{equation}
	\mathbf{K}^{(M)}_s (k+3, k) = \mathbf{K}^{(M)}_s (k, k+3)  = 	\sin^2(k\pi/2) c_1^2\nu_2^4    \ \ \ \ k = 1, 2, ..., M-3 
	\end{equation}

\section{Result for replacing the second OPA of the unbalanced SU(1,1) interferometer with a 50:50 beam-splitter}

\subsection{the covariance matrix result for $\rho_{bs}^{(M)}$}

Replace OPA2 with a 50:50 beam splitter in an unbalanced SU(1,1) interferometer, the CM of the displacement vector for a Gaussian state $\rho_{bs}^{(M)}$ generated from this interferometer pumping by finite number of pump pulses $M-1$ is defined as
\begin{equation}
	\label{CM_bs_def}
	\sigma_{c,bs}^{(M)} = \left[\begin{matrix} \mathbf{A}_{bs}^{(M)} & \mathbf{B}_{bs}^{(M)} \\ \mathbf{ B}_{bs}^{(M)*} & \mathbf{ A}_{bs}^{(M)*} \end{matrix}\right].
\end{equation}
The calculated $\mathbf{A}_{bs}^{(M)}$ and $\mathbf{B}_{bs}^{(M)}$ for 
$M=2$, $M=3$ and $M=4$ are given by
\begin{align}
	\mathbf{A}_{bs}^{(2)} =  \left[\begin{matrix}1 & 0\\0 & 1\end{matrix}\right], 
	~~~
	\mathbf{B}_{bs}^{(2)} = \left[\begin{matrix}0 & 0\\0 & 0\end{matrix}\right],
	\end{align}
\begin{align}
	\mathbf{A}_{bs}^{(3)} =  \left[\begin{matrix}\frac{\mu_{1}^{2}}{2} + \frac{\nu_{1}^{2}}{2} + \frac{1}{2} & - \frac{\mu_{1}^{2}}{2} - \frac{\nu_{1}^{2}}{2} + \frac{1}{2} & 0\\- \frac{\mu_{1}^{2}}{2} - \frac{\nu_{1}^{2}}{2} + \frac{1}{2} & \frac{\mu_{1}^{2}}{2} + \frac{\nu_{1}^{2}}{2} + \frac{1}{2} & 0\\0 & 0 & V_{1}\end{matrix}\right],
	~~~
	\mathbf{B}_{bs}^{(3)} = \left[\begin{matrix}0 & 0 & - \sqrt{2} c_{1}\\0 & 0 & \sqrt{2} c_{1}\\- \sqrt{2} c_{1} & \sqrt{2} c_{1} & 0\end{matrix}\right],
\end{align}
and
\begin{equation}
	\mathbf{A}_{bs}^{(4)} = \left[\begin{matrix}\frac{\mu_{1}^{2}}{2} + \frac{\nu_{1}^{2}}{2} + \frac{1}{2} & - \frac{\mu_{1}^{2}}{2} - \frac{\nu_{1}^{2}}{2} + \frac{1}{2} & 0 & 0\\- \frac{\mu_{1}^{2}}{2} - \frac{\nu_{1}^{2}}{2} + \frac{1}{2} & \frac{\mu_{1}^{2}}{2} + \frac{\nu_{1}^{2}}{2} + \frac{1}{2} & 0 & 0\\0 & 0 & \frac{\mu_{1}^{2}}{2} + \frac{\nu_{1}^{2}}{2} + \frac{1}{2} & \frac{\mu_{1}^{2}}{2} + \frac{\nu_{1}^{2}}{2} - \frac{1}{2}\\0 & 0 & \frac{\mu_{1}^{2}}{2} + \frac{\nu_{1}^{2}}{2} - \frac{1}{2} & \frac{\mu_{1}^{2}}{2} + \frac{\nu_{1}^{2}}{2} + \frac{1}{2}\end{matrix}\right]
, ~~~
	\mathbf{B}_{bs}^{(4)} = \left[\begin{matrix}0 & 0 & - c_{1} e^{i \phi} & - c_{1} e^{i \phi}\\0 & 0 & c_{1} e^{i \phi} & c_{1} e^{i \phi}\\- c_{1} e^{i \phi} & c_{1} e^{i \phi} & 0 & 0\\- c_{1} e^{i \phi} & c_{1} e^{i \phi} & 0 & 0\end{matrix}\right],
\end{equation}
respectively. For $M=5$ and $M=6$, 
\begin{align}
	\mathbf{A}_{bs}^{(5)} =& \left[\begin{matrix}\frac{\mu_{1}^{2}}{2} + \frac{\nu_{1}^{2}}{2} + \frac{1}{2} & - \frac{\mu_{1}^{2}}{2} - \frac{\nu_{1}^{2}}{2} + \frac{1}{2} & 0 & 0 & 0\\- \frac{\mu_{1}^{2}}{2} - \frac{\nu_{1}^{2}}{2} + \frac{1}{2} & \frac{\mu_{1}^{2}}{2} + \frac{\nu_{1}^{2}}{2} + \frac{1}{2} & 0 & 0 & 0\\0 & 0 & V_{1} & 0 & 0\\0 & 0 & 0 & V_{1} & 0\\0 & 0 & 0 & 0 & V_{1}\end{matrix}\right], 
~~~
	\mathbf{B}_{bs}^{(5)} = \left[\begin{matrix}0 & 0 & - c_{1} e^{i \phi} & - c_{1} e^{i \phi} & 0\\0 & 0 & c_{1} e^{i \phi} & c_{1} e^{i \phi} & 0\\- c_{1} e^{i \phi} & c_{1} e^{i \phi} & 0 & 0 & - \sqrt{2} c_{1}\\- c_{1} e^{i \phi} & c_{1} e^{i \phi} & 0 & 0 & \sqrt{2} c_{1}\\0 & 0 & - \sqrt{2} c_{1} & \sqrt{2} c_{1} & 0\end{matrix}\right],
\end{align}
and
\begin{align}
	\mathbf{A}_{bs}^{(6)} = \left[\begin{matrix}\frac{\mu_{1}^{2}}{2} + \frac{\nu_{1}^{2}}{2} + \frac{1}{2} & - \frac{\mu_{1}^{2}}{2} - \frac{\nu_{1}^{2}}{2} + \frac{1}{2} & 0 & 0 & 0 & 0\\- \frac{\mu_{1}^{2}}{2} - \frac{\nu_{1}^{2}}{2} + \frac{1}{2} & \frac{\mu_{1}^{2}}{2} + \frac{\nu_{1}^{2}}{2} + \frac{1}{2} & 0 & 0 & 0 & 0\\0 & 0 & V_{1} & 0 & 0 & 0\\0 & 0 & 0 & V_{1} & 0 & 0\\0 & 0 & 0 & 0 & \frac{\mu_{1}^{2}}{2} + \frac{\nu_{1}^{2}}{2} + \frac{1}{2} & \frac{\mu_{1}^{2}}{2} + \frac{\nu_{1}^{2}}{2} - \frac{1}{2}\\0 & 0 & 0 & 0 & \frac{\mu_{1}^{2}}{2} + \frac{\nu_{1}^{2}}{2} - \frac{1}{2} & \frac{\mu_{1}^{2}}{2} + \frac{\nu_{1}^{2}}{2} + \frac{1}{2}\end{matrix}\right],
\end{align}
\begin{align}
	\mathbf{B}_{bs}^{(6)} = \left[\begin{matrix}0 & 0 & - c_{1} e^{i \phi} & - c_{1} e^{i \phi} & 0 & 0\\0 & 0 & c_{1} e^{i \phi} & c_{1} e^{i \phi} & 0 & 0\\- c_{1} e^{i \phi} & c_{1} e^{i \phi} & 0 & 0 & - c_{1} e^{i \phi} & - c_{1} e^{i \phi}\\- c_{1} e^{i \phi} & c_{1} e^{i \phi} & 0 & 0 & c_{1} e^{i \phi} & c_{1} e^{i \phi}\\0 & 0 & - c_{1} e^{i \phi} & c_{1} e^{i \phi} & 0 & 0\\0 & 0 & - c_{1} e^{i \phi} & c_{1} e^{i \phi} & 0 & 0\end{matrix}\right].
\end{align}

The analytical result of the non-zero elements of $\mathbf{A}_{bs}^{(M)}$ for odd number $M ~(M>4)$ are given by
\begin{align}
	\mathbf{A}_{bs}^{(M)} (k, k)  
	= &
	 \left\{ 
	\begin{array}{cl}
		\frac{\mu_{1}^{2}}{2} + \frac{\nu_{1}^{2}}{2} + \frac{1}{2} 	&  \ \ \ \ k = 1 ~or ~k=2 \\
		V_1                   &  \ \ \ \ k = 3, 4, ..., M   \\
	\end{array}\right. \\
     \mathbf{A}_{bs}^{(M)} (k, k+1)
     = & \mathbf{A}_{bs}^{(M)} (k+1, k)
     = - \frac{\mu_{1}^{2}}{2} - \frac{\nu_{1}^{2}}{2} + \frac{1}{2}
     \ \ \ \ k=1
\end{align}

The analytical result of the non-zero elements of $\mathbf{A}_{bs}^{(M)}$ for even number $M ~(M>4)$ are given by
\begin{align}
	\mathbf{A}_{bs}^{(M)} (k, k)  
	= & 
	\left\{ 
	\begin{array}{cl}
		\frac{\mu_{1}^{2}}{2} + \frac{\nu_{1}^{2}}{2} + \frac{1}{2} 	&  \ \ \ \ k = 1 ~or ~k=2 ~or ~k=M-1 ~or ~k=M \\
		V_1                   &  \ \ \ \ k = 3, 4, ..., M-2   
	\end{array}\right. 
 \\
     \mathbf{A}_{bs}^{(M)} (k, k+1)
     	=  &
     	\mathbf{A}_{bs}^{(M)} (k+1, k) =
     \left\{
     \begin{array}{cl}
        - \frac{\mu_{1}^{2}}{2} - \frac{\nu_{1}^{2}}{2} + \frac{1}{2}
        &  \ \ \ \ k=1 \\
        \frac{\mu_{1}^{2}}{2} + \frac{\nu_{1}^{2}}{2} - \frac{1}{2} 
         &  \ \ \ \ k=M-1 
     \end{array}\right. 
\end{align}

For odd number $M ~(M>4)$, the non-zero elements of $\mathbf{B}_{bs}^{(M)}$ can be expressed as
\begin{align}
	\mathbf{B}_{bs}^{(M)} (k, k+1) = & \mathbf{B}_{bs}^{(M)} (k+1, k)=  
	\left\{ 
	\begin{array}{cl}
	\cos^2(k\pi/2) c_{1} e^{i \phi}          &\ \ \ \ k = 1, 2, ..., M-2
	\\
	\sqrt{2} c_{1}            &\ \ \ \ k = M-1
    \end{array}\right. 
\\
    \mathbf{B}_{bs}^{(M)} (k, k+2) = & \mathbf{B}_{bs}^{(M)} (k+2, k)=  
    \left\{ 
    \begin{array}{cl}
    	(-1)^{k} c_{1} e^{i \phi}          &\ \ \ \ k = 1, 2, ..., M-3
    	\\
    	-\sqrt{2} c_{1}            &\ \ \ \ k = M-2
    \end{array}\right. 
\\
    \mathbf{B}_{bs}^{(M)} (k, k+3) = & \mathbf{B}_{bs}^{(M)} (k+3, k)= 
    -\sin^2(k\pi/2) c_{1} e^{i \phi}             \ \ \ \ k = 1, 2, ..., M-3
\end{align}

For even number $M ~(M>4)$, the non-zero elements of $\mathbf{B}_{bs}^{(M)}$ can be expressed as
\begin{align}
	\mathbf{B}_{bs}^{(M)} (k, k+1) = & \mathbf{B}_{bs}^{(M)} (k+1, k)=  
	\cos^2(k\pi/2)	c_{1} e^{i \phi}          \ \ \ \ k = 1, 2, ..., M-1
	\\
	\mathbf{B}_{bs}^{(M)} (k, k+2) = & \mathbf{B}_{bs}^{(M)} (k+2, k)=  
		(-1)^{k} c_{1} e^{i \phi}          \ \ \ \ k = 1, 2, ..., M-2
	\\
	\mathbf{B}_{bs}^{(M)} (k, k+3) = & \mathbf{B}_{bs}^{(M)} (k+3, k)= 
	- \sin^2(k\pi/2) c_{1} e^{i \phi}             \ \ \ \ k = 1, 2, ..., M-3
\end{align}

\subsection{the covariance matrix result for $\rho_{bs,s}^{(M)}$}

We define the density operator of $M$-mode output state from an interferometer, of which OPA2 in an unbalanced SU(1,1) interferometer is replaced with a 50:50 beam splitter, pumping by a stable pulse train as $\rho_{bs,s}^{M}$. Similar to the unbalanced SU(1,1) interferomter, the state $\rho_{bs,s}^{M}$ can be viewed as a subsystem of $\rho_{bs}^{M+4}$. For $M=2$, $M=3$ and $M=4$, the calculated block matrixs $\mathbf{A}_{bs,s}^{(M)}$ and $\mathbf{B}_{bs,s}^{(M)}$ of the CM $\sigma_{c,bs,s}^{(M)}$ for $\rho_{bs,s}^{M}$ are given by
\begin{align}
	\mathbf{A}_{bs,s}^{(2)} = \left[\begin{matrix}V_{1} & 0\\0 & V_{1}\end{matrix}\right],
	~~~
	\mathbf{B}_{bs,s}^{(2)} =\left[\begin{matrix}0 & 0\\0 & 0\end{matrix}\right], 
	\end{align}
\begin{align}
	\mathbf{A}_{bs,s}^{(3)} = \left[\begin{matrix}V_{1} & 0 & 0\\0 & V_{1} & 0\\0 & 0 & V_{1}\end{matrix}\right] ,
	~~~
	\mathbf{B}_{bs,s}^{(3)} = \left[\begin{matrix}0 & 0 & - c_{1} e^{i \phi}\\0 & 0 & c_{1} e^{i \phi}\\- c_{1} e^{i \phi} & c_{1} e^{i \phi} & 0\end{matrix}\right] , 
\end{align}
and
\begin{align}
	\mathbf{A}_{bs,s}^{(4)} =  \left[\begin{matrix}V_{1} & 0 & 0 & 0\\0 & V_{1} & 0 & 0\\0 & 0 & V_{1} & 0\\0 & 0 & 0 & V_{1}\end{matrix}\right] ,
	~~~
	\mathbf{B}_{bs,s}^{(4)} =  \left[\begin{matrix}0 & 0 & - c_{1} e^{i \phi} & - c_{1} e^{i \phi}\\0 & 0 & c_{1} e^{i \phi} & c_{1} e^{i \phi}\\- c_{1} e^{i \phi} & c_{1} e^{i \phi} & 0 & 0\\- c_{1} e^{i \phi} & c_{1} e^{i \phi} & 0 & 0\end{matrix}\right] . 
\end{align}

The non-zero elements of the block matrix $\mathbf{A}_{bs,s}^{(M)}$ with mode number $M ~(M>3)$ can be expressed as
\begin{align}
	\mathbf{A}_{bs,s}^{(M)}(k,k) = V_1            \ \ \ \ k = 1, 2, ..., M
	\end{align}

The non-zero elements of the block matrix $\mathbf{B}_{bs,s}^{(M)}$ with mode number $M ~(M>3)$ can be expressed as
\begin{align}
	\mathbf{B}_{bs,s}^{(M)} (k, k+1) = & \mathbf{B}_{bs,s}^{(M)} (k+1, k)=  
	\cos^2(k\pi/2) c_{1} e^{i \phi}          \ \ \ \ k = 1, 2, ..., M-1
	\\
	\mathbf{B}_{bs,s}^{(M)} (k, k+2) = & \mathbf{B}_{bs,s}^{(M)} (k+2, k)=  
	(-1)^{k} c_{1} e^{i \phi}          \ \ \ \ k = 1, 2, ..., M-2
	\\
	\mathbf{B}_{bs,s}^{(M)} (k, k+3) = & \mathbf{B}_{bs,s}^{(M)} (k+3, k)= 
	- \sin^2(k\pi/2) c_{1} e^{i \phi}             \ \ \ \ k = 1, 2, ..., M-3
\end{align}

\subsection{the photon number statistics property for $\rho_{bs}^{(M)}$}

The average photon number vector $\mathbf{m}^{(M)}_{bs}$ of the state $\rho^{(M)}_{bs}$ for $M=2$, $M=3$, and $M=4$ are
\begin{equation}
  \mathbf{m}^{(2)}_{bs} = \left[\begin{matrix}0 & 0\end{matrix}\right]^T,
\end{equation}
\begin{equation}
    \mathbf{m}^{(3)}_{bs} =   \left[\begin{matrix}0.5 \nu_{1}^{2} & 0.5 \nu_{1}^{2} & \nu_{1}^{2}\end{matrix}\right]^T,
\end{equation}
and
\begin{equation}
   \mathbf{m}^{(4)}_{bs} =  \left[\begin{matrix}0.5 \nu_{1}^{2} & 0.5 \nu_{1}^{2} & 0.5 \nu_{1}^{2} & 0.5 \nu_{1}^{2}\end{matrix}\right]^T   ,
\end{equation}
respectively. For $M=5$ and $M=6$,
\begin{equation}
    \mathbf{m}^{(5)}_{bs} =  \left[\begin{matrix}0.5 \nu_{1}^{2} & 0.5 \nu_{1}^{2} & \nu_{1}^{2} & \nu_{1}^{2} & \nu_{1}^{2}\end{matrix}\right]^T,
\end{equation}
and
\begin{equation}
 \mathbf{m}^{(6)}_{bs} =  \left[\begin{matrix}0.5 \nu_{1}^{2} & 0.5 \nu_{1}^{2} & \nu_{1}^{2} & \nu_{1}^{2} & 0.5 \nu_{1}^{2} & 0.5 \nu_{1}^{2}\end{matrix}\right] ^T .  
\end{equation}
The analytical result of the average photon number vector $\mathbf{m}^{(M)}_{bs}$ for an odd number $M ~(M>4)$ has the form of 
\begin{align}
\mathbf{m}^{(M)}_{bs} = \Big[\frac{1}{2}\nu_{1}^{2} ~~ \frac{1}{2}\nu_{1}^{2} ~~\underbrace{\nu_{1}^{2} ~~\nu_{1}^{2} ~~ \cdots }_{ M - 2}  \Big]^T,
\end{align}
The analytical result of the average photon number vector $\mathbf{m}^{(M)}_{bs}$ for an even number $M ~(M>4)$ has the form of 
\begin{align}
\mathbf{m}^{(M)}_{bs} = \Big[\frac{1}{2}\nu_{1}^{2} ~~ \frac{1}{2}\nu_{1}^{2} ~~\underbrace{\nu_{1}^{2} ~~\nu_{1}^{2} ~~ \cdots }_{ M - 4} ~~ \frac{1}{2}\nu_{1}^{2} ~~ \frac{1}{2}\nu_{1}^{2} \Big]^T.
\end{align}

For $M=2$, $M=3$ and $M=4$, the CM $\mathbf{K}^{(M)}_{bs}$ of the state $\rho_{bs}^{(M)}$ for the photon number operator are
\begin{equation}
   \mathbf{K}^{(2)}_{bs} = \left[\begin{matrix}0 & 0\\0 & 0\end{matrix}\right],
\end{equation}
\begin{equation}
   \mathbf{K}^{(3)}_{bs} = \left[\begin{matrix}\frac{1}{4} \mu_{1}^{4} - \frac{1}{4} & \frac{1}{4} \nu_{1}^{4} & \frac{1}{4} \mu_{1}^{4} + \frac{1}{4} \nu_{1}^{4} - \frac{1}{4}\\\frac{1}{4} \nu_{1}^{4} & \frac{1}{4} \mu_{1}^{4} - \frac{1}{4} & \frac{1}{4} \mu_{1}^{4} + \frac{1}{4} \nu_{1}^{4} - \frac{1}{4}\\\frac{1}{4} \mu_{1}^{4} + \frac{1}{4} \nu_{1}^{4} - \frac{1}{4} & \frac{1}{4} \mu_{1}^{4} + \frac{1}{4} \nu_{1}^{4} - \frac{1}{4} & \frac{1}{2} \mu_{1}^{4} + \frac{1}{2} \nu_{1}^{4} - \frac{1}{2}\end{matrix}\right],
\end{equation}
and
\begin{equation}
    \mathbf{K}^{(4)}_{bs} = \left[\begin{matrix}\frac{1}{4} \mu_{1}^{4} - \frac{1}{4} & \frac{1}{4} \nu_{1}^{4} & \frac{1}{8} \mu_{1}^{4} + \frac{1}{8} \nu_{1}^{4} - \frac{1}{8} & \frac{1}{8} \mu_{1}^{4} +\frac{1}{8} \nu_{1}^{4} - \frac{1}{8}\\\frac{1}{4} \nu_{1}^{4} & \frac{1}{4} \mu_{1}^{4} - \frac{1}{4} & \frac{1}{8} \mu_{1}^{4} + \frac{1}{8} \nu_{1}^{4} - \frac{1}{8} & \frac{1}{8} \mu_{1}^{4} + \frac{1}{8} \nu_{1}^{4} -\frac{1}{8}\\\frac{1}{8} \mu_{1}^{4} + \frac{1}{8} \nu_{1}^{4} - \frac{1}{8}& \frac{1}{8} \mu_{1}^{4} + \frac{1}{8} \nu_{1}^{4} - \frac{1}{8} & \frac{1}{4} \mu_{1}^{4} - \frac{1}{4} & \frac{1}{4} \nu_{1}^{4}\\\frac{1}{8} \mu_{1}^{4} + \frac{1}{8} \nu_{1}^{4} - \frac{1}{8} & \frac{1}{8} \mu_{1}^{4} + \frac{1}{8} \nu_{1}^{4} - \frac{1}{8} & \frac{1}{4} \nu_{1}^{4} & \frac{1}{4} \mu_{1}^{4} - \frac{1}{4}\end{matrix}\right],
\end{equation}
respectively. For an odd number $M>4$, the non-zero elements of the CM $\mathbf{K}^{(M)}_{bs}$ are
\begin{equation}
	\mathbf{K}^{(M)}_{bs} (k, k)  = \left\{ 
	\begin{array}{cl}
	    (\mu_{1}^{4} - 1)/4	&  \ \ \ \ k = 1 ~or ~ k = 2\\
		 (\mu_{1}^{4} +  \nu_{1}^{4} - 1)/2    &  \ \ \ \ k = 3, 4, ..., M 
	\end{array}\right.
\end{equation}
\begin{equation}
	\mathbf{K}^{(M)}_{bs} (k+1, k) = \mathbf{K}^{(M)}_{bs} (k, k+1)  = \left\{ 
	\begin{array}{cl}
	    \nu_1^4/4 	&  \ \ \ \ k = 1 \\
		\cos^2(k\pi/2) ( \mu_{1}^{4} +  \nu_{1}^{4} - 1)/8  &  \ \ \ \ k = 2, 3, ..., M-2  \\
		 ( \mu_{1}^{4} +  \nu_{1}^{4} - 1)/4  &  \ \ \ \ k = M-1  
	\end{array}\right.
\end{equation}
\begin{equation}
	\mathbf{K}^{(M)}_{bs} (k+2, k) = \mathbf{K}^{(M)}_{bs} (k, k+2)  = \left\{ 
	\begin{array}{cl}
	( \mu_{1}^{4} +  \nu_{1}^{4} - 1)/8  &  \ \ \ \ k = 1, 2, ..., M-3  \\
		 ( \mu_{1}^{4} +  \nu_{1}^{4} - 1)/4  &  \ \ \ \ k = M-2  
	\end{array}\right.
\end{equation}
\begin{equation}
	\mathbf{K}^{(M)}_{bs} (k+3, k) = \mathbf{K}^{(M)}_{bs} (k, k+3)  = 
			\sin^2(k\pi/2) ( \mu_{1}^{4} +  \nu_{1}^{4} - 1)/8    \ \ \ \ k = 1, 2, ..., M-3   
	\end{equation}

For an even number $M>4$, the non-zero elements of the CM $\mathbf{K}^{(M)}_{bs}$ are
\begin{equation}
	\mathbf{K}^{(M)}_{bs} (k, k)  = \left\{ 
	\begin{array}{cl}
	    (\mu_{1}^{4} - 1)/4	&  \ \ \ \ k = 1 ~or ~ k = 2 ~or ~k = M-1 ~or ~k = M\\
		 (\mu_{1}^{4} +  \nu_{1}^{4} - 1)/2    &  \ \ \ \ k = 3, 4, ..., M-2 
	\end{array}\right.
\end{equation}
\begin{equation}
	\mathbf{K}^{(M)}_{bs} (k+1, k) = \mathbf{K}^{(M)}_{bs} (k, k+1)  = \left\{ 
	\begin{array}{cl}
	    \nu_1^4/4 	&  \ \ \ \ k = 1 ~or ~ k = M-1 \\
		\cos^2(k\pi/2) ( \mu_{1}^{4} +  \nu_{1}^{4} - 1)/8  &  \ \ \ \ k = 2, 3, ..., M-2  
	\end{array}\right.
\end{equation}
\begin{equation}
	\mathbf{K}^{(M)}_{bs} (k+2, k) = \mathbf{K}^{(M)}_{bs} (k, k+2)  = 
	( \mu_{1}^{4} +  \nu_{1}^{4} - 1)/8    \ \ \ \ k = 1, 2, ..., M-2  
\end{equation}
\begin{equation}
	\mathbf{K}^{(M)}_{bs} (k+3, k) = \mathbf{K}^{(M)}_{bs} (k, k+3)  = \sin^2(k\pi/2) ( \mu_{1}^{4} +  \nu_{1}^{4} - 1)/8    \ \ \ \ k = 1, 2, ..., M-3   
	\end{equation}


\subsection{the photon number statistics property for $\rho_{bs,s}^{(M)}$}

The average photon number vector $\mathbf{m}^{(M)}_{bs,s}$ of the state $\rho^{(M)}_{bs,s}$ with mode number $M ~(M\geq 2)$ has the form of 
\begin{align}
\mathbf{m}^{(M)}_{bs,s} = \Big[ \underbrace{\nu_{1}^{2} ~~\nu_{1}^{2} ~~ \cdots }_{ M }  \Big]^T.
\end{align}

The non-zero elements of the CM $\mathbf{K}^{(M)}_{bs,s}$  of the state $\rho^{(M)}_{bs,s}$ with mode number $M=2$ and $M=3$ are
\begin{equation}
\mathbf{K}^{(2)}_{bs,s} = \left[\begin{matrix}\frac{1}{2} \mu_{1}^{4} + \frac{1}{2} \nu_{1}^{4} - \frac{1}{2} & 0\\0 & \frac{1}{2} \mu_{1}^{4} + \frac{1}{2} \nu_{1}^{4} - \frac{1}{2}\end{matrix}\right] ,   
\end{equation}
and
\begin{equation}
\mathbf{K}^{(3)}_{bs,s} = \left[\begin{matrix}\frac{1}{2} \mu_{1}^{4} + \frac{1}{2} \nu_{1}^{4} - \frac{1}{2} & 0 & \frac{1}{8} \mu_{1}^{4} + \frac{1}{8} \nu_{1}^{4} - \frac{1}{8}\\0 & \frac{1}{2} \mu_{1}^{4} + \frac{1}{2} \nu_{1}^{4} -\frac{1}{2} & \frac{1}{8} \mu_{1}^{4} + \frac{1}{8} \nu_{1}^{4} - \frac{1}{8}\\\frac{1}{8} \mu_{1}^{4} + \frac{1}{8} \nu_{1}^{4} - \frac{1}{8} & \frac{1}{8} \mu_{1}^{4} + \frac{1}{8} \nu_{1}^{4} -\frac{1}{8} & \frac{1}{2} \mu_{1}^{4} + \frac{1}{2} \nu_{1}^{4} - \frac{1}{2}\end{matrix}\right],
\end{equation}
respectively.
The non-zero elements of the CM $\mathbf{K}^{(M)}_{bs,s}$  of the state $\rho^{(M)}_{bs,s}$ with mode number $M ~(M > 3)$ are
\begin{equation}
	\mathbf{K}^{(M)}_{bs,s} (k, k)  = 
		 (\mu_{1}^{4} +  \nu_{1}^{4} - 1)/2      \ \ \ \ k = 1, 2, ..., M 
\end{equation}
\begin{equation}
	\mathbf{K}^{(M)}_{bs,s} (k+1, k) = \mathbf{K}^{(M)}_{bs,s} (k, k+1)  =
		\cos^2(k\pi/2) ( \mu_{1}^{4} +  \nu_{1}^{4} - 1)/8    \ \ \ \ k = 1, 2, ..., M-1  
\end{equation}
\begin{equation}
	\mathbf{K}^{(M)}_{bs,s} (k+2, k) = \mathbf{K}^{(M)}_{bs,s} (k, k+2)  = 
	( \mu_{1}^{4} +  \nu_{1}^{4} - 1)/8    \ \ \ \ k = 1, 2, ..., M-2  
\end{equation}
\begin{equation}
	\mathbf{K}^{(M)}_{bs,s} (k+3, k) = \mathbf{K}^{(M)}_{bs,s} (k, k+3)  = \sin^2(k\pi/2) ( \mu_{1}^{4} +  \nu_{1}^{4} - 1)/8    \ \ \ \ k = 1, 2, ..., M-3   
	\end{equation}

\section{PPT Negativity of all bi-partite formed using modes in the state of $\hat \rho_s^{(6)}$}

We use the notation of A and B to represent the bi-partite formed using modes in the state of $\hat \rho_s^{(6)}$. When A and B are a bi-partite divisions of $\hat \rho_s^{(6)}$, which means $A \cup B = \{ 1, 2, 3, 4, 5, 6 \}$, the PPT negativity is proven. Here we list the case of $A \cup B \neq \{ 1, 2, 3, 4, 5, 6 \}$ with the following 5 tables. We divide the characteristics of PPT Negativity into 3 catalogs: (1) always exist: the minimum PPT eign value is always negative for any non-zero $r_1$ and $r_2$ value; (2) partially exist: the minimum PPT eign value is negative only in some of the non-zero value of $r_1$ and $r_2$; (3) None: the minimum PPT eign value is not negative for any $r_1$ and $r_2$ value.

\begin{table}[h]
\begin{tabular}{||p{1cm}|p{1.2cm}|p{2cm}||p{1cm}|p{1.2cm}|p{2cm}||p{1cm}|p{1.2cm}|p{2cm}||}
\cline{1-9}
A               &   B           &   PPT Neg.     &   A          &    B           &   PPT Neg.        &  A           &    B           &   PPT Neg.  \\
\cline{1-9}
$\{$1$\}$ & $\{$2$\}$ & partially exist &$\{$1$\}$ & $\{$3$\}$ & none &$\{$1$\}$ & $\{$4$\}$ & none  \\ 
$\{$1$\}$ & $\{$5$\}$ & none &$\{$1$\}$ & $\{$6$\}$ & none &$\{$2$\}$ & $\{$3$\}$ & partially exist  \\ 
$\{$2$\}$ & $\{$4$\}$ & none &$\{$2$\}$ & $\{$5$\}$ & none &$\{$2$\}$ & $\{$6$\}$ & none  \\ 
$\{$3$\}$ & $\{$4$\}$ & partially exist &$\{$3$\}$ & $\{$5$\}$ & none &$\{$3$\}$ & $\{$6$\}$ & none  \\ 
$\{$4$\}$ & $\{$5$\}$ & partially exist &$\{$4$\}$ & $\{$6$\}$ & none &$\{$5$\}$ & $\{$6$\}$ & partially exist  \\ 
\cline{1-9}
\end{tabular}
\end{table}

\begin{table}[h]
\begin{tabular}{||p{1cm}|p{1.2cm}|p{2cm}||p{1cm}|p{1.2cm}|p{2cm}||p{1cm}|p{1.2cm}|p{2cm}||}
\cline{1-9}
A               &   B           &   PPT Neg.     &   A          &    B           &   PPT Neg.        &  A           &    B           &   PPT Neg.  \\
\cline{1-9}
$\{$1$\}$ & $\{$2, 3$\}$ & partially exist &$\{$1$\}$ & $\{$2, 4$\}$ & partially exist &$\{$1$\}$ & $\{$2, 5$\}$ & partially exist  \\ 
$\{$1$\}$ & $\{$2, 6$\}$ & partially exist &$\{$1$\}$ & $\{$3, 4$\}$ & none &$\{$1$\}$ & $\{$3, 5$\}$ & none  \\ 
$\{$1$\}$ & $\{$3, 6$\}$ & none &$\{$1$\}$ & $\{$4, 5$\}$ & none &$\{$1$\}$ & $\{$4, 6$\}$ & none  \\ 
$\{$1$\}$ & $\{$5, 6$\}$ & none &$\{$2$\}$ & $\{$1, 3$\}$ & partially exist &$\{$2$\}$ & $\{$1, 4$\}$ & partially exist  \\ 
$\{$2$\}$ & $\{$1, 5$\}$ & partially exist &$\{$2$\}$ & $\{$1, 6$\}$ & partially exist &$\{$2$\}$ & $\{$3, 4$\}$ & partially exist  \\ 
$\{$2$\}$ & $\{$3, 5$\}$ & partially exist &$\{$2$\}$ & $\{$3, 6$\}$ & partially exist &$\{$2$\}$ & $\{$4, 5$\}$ & none  \\ 
$\{$2$\}$ & $\{$4, 6$\}$ & none &$\{$2$\}$ & $\{$5, 6$\}$ & none &$\{$3$\}$ & $\{$1, 2$\}$ & partially exist  \\ 
$\{$3$\}$ & $\{$1, 4$\}$ & partially exist &$\{$3$\}$ & $\{$1, 5$\}$ & none &$\{$3$\}$ & $\{$1, 6$\}$ & none  \\ 
$\{$3$\}$ & $\{$2, 4$\}$ & partially exist &$\{$3$\}$ & $\{$2, 5$\}$ & partially exist &$\{$3$\}$ & $\{$2, 6$\}$ & partially exist  \\ 
$\{$3$\}$ & $\{$4, 5$\}$ & partially exist &$\{$3$\}$ & $\{$4, 6$\}$ & partially exist &$\{$3$\}$ & $\{$5, 6$\}$ & none  \\ 
$\{$4$\}$ & $\{$1, 2$\}$ & none &$\{$4$\}$ & $\{$1, 3$\}$ & partially exist &$\{$4$\}$ & $\{$1, 5$\}$ & partially exist  \\ 
$\{$4$\}$ & $\{$1, 6$\}$ & none &$\{$4$\}$ & $\{$2, 3$\}$ & partially exist &$\{$4$\}$ & $\{$2, 5$\}$ & partially exist  \\ 
$\{$4$\}$ & $\{$2, 6$\}$ & none &$\{$4$\}$ & $\{$3, 5$\}$ & partially exist &$\{$4$\}$ & $\{$3, 6$\}$ & partially exist  \\ 
$\{$4$\}$ & $\{$5, 6$\}$ & partially exist &$\{$5$\}$ & $\{$1, 2$\}$ & none &$\{$5$\}$ & $\{$1, 3$\}$ & none  \\ 
$\{$5$\}$ & $\{$1, 4$\}$ & partially exist &$\{$5$\}$ & $\{$1, 6$\}$ & partially exist &$\{$5$\}$ & $\{$2, 3$\}$ & none  \\ 
$\{$5$\}$ & $\{$2, 4$\}$ & partially exist &$\{$5$\}$ & $\{$2, 6$\}$ & partially exist &$\{$5$\}$ & $\{$3, 4$\}$ & partially exist  \\ 
$\{$5$\}$ & $\{$3, 6$\}$ & partially exist &$\{$5$\}$ & $\{$4, 6$\}$ & partially exist &$\{$6$\}$ & $\{$1, 2$\}$ & none  \\ 
$\{$6$\}$ & $\{$1, 3$\}$ & none &$\{$6$\}$ & $\{$1, 4$\}$ & none &$\{$6$\}$ & $\{$1, 5$\}$ & partially exist  \\ 
$\{$6$\}$ & $\{$2, 3$\}$ & none &$\{$6$\}$ & $\{$2, 4$\}$ & none &$\{$6$\}$ & $\{$2, 5$\}$ & partially exist  \\ 
$\{$6$\}$ & $\{$3, 4$\}$ & none &$\{$6$\}$ & $\{$3, 5$\}$ & partially exist &$\{$6$\}$ & $\{$4, 5$\}$ & partially exist  \\ 
\cline{1-9}
\end{tabular}
\end{table}

\begin{table}[h]
\begin{tabular}{||p{1cm}|p{1.2cm}|p{2cm}||p{1cm}|p{1.2cm}|p{2cm}||p{1cm}|p{1.2cm}|p{2cm}||}
\cline{1-9}
A               &   B           &   PPT Neg.     &   A          &    B           &   PPT Neg.        &  A           &    B           &   PPT Neg.  \\
\cline{1-9}
$\{$1$\}$ & $\{$2, 3, 4$\}$ & always exist &$\{$1$\}$ & $\{$2, 3, 5$\}$ & partially exist &$\{$1$\}$ & $\{$2, 3, 6$\}$ & partially exist  \\ 
$\{$1$\}$ & $\{$2, 4, 5$\}$ & partially exist &$\{$1$\}$ & $\{$2, 4, 6$\}$ & partially exist &$\{$1$\}$ & $\{$2, 5, 6$\}$ & partially exist  \\ 
$\{$1$\}$ & $\{$3, 4, 5$\}$ & none &$\{$1$\}$ & $\{$3, 4, 6$\}$ & none &$\{$1$\}$ & $\{$3, 5, 6$\}$ & none  \\ 
$\{$1$\}$ & $\{$4, 5, 6$\}$ & none &$\{$2$\}$ & $\{$1, 3, 4$\}$ & always exist &$\{$2$\}$ & $\{$1, 3, 5$\}$ & partially exist  \\ 
$\{$2$\}$ & $\{$1, 3, 6$\}$ & partially exist &$\{$2$\}$ & $\{$1, 4, 5$\}$ & partially exist &$\{$2$\}$ & $\{$1, 4, 6$\}$ & partially exist  \\ 
$\{$2$\}$ & $\{$1, 5, 6$\}$ & partially exist &$\{$2$\}$ & $\{$3, 4, 5$\}$ & partially exist &$\{$2$\}$ & $\{$3, 4, 6$\}$ & partially exist  \\ 
$\{$2$\}$ & $\{$3, 5, 6$\}$ & partially exist &$\{$2$\}$ & $\{$4, 5, 6$\}$ & none &$\{$3$\}$ & $\{$1, 2, 4$\}$ & always exist  \\ 
$\{$3$\}$ & $\{$1, 2, 5$\}$ & partially exist &$\{$3$\}$ & $\{$1, 2, 6$\}$ & partially exist &$\{$3$\}$ & $\{$1, 4, 5$\}$ & partially exist  \\ 
$\{$3$\}$ & $\{$1, 4, 6$\}$ & partially exist &$\{$3$\}$ & $\{$1, 5, 6$\}$ & none &$\{$3$\}$ & $\{$2, 4, 5$\}$ & partially exist  \\ 
$\{$3$\}$ & $\{$2, 4, 6$\}$ & partially exist &$\{$3$\}$ & $\{$2, 5, 6$\}$ & partially exist &$\{$3$\}$ & $\{$4, 5, 6$\}$ & always exist  \\ 
$\{$4$\}$ & $\{$1, 2, 3$\}$ & always exist &$\{$4$\}$ & $\{$1, 2, 5$\}$ & partially exist &$\{$4$\}$ & $\{$1, 2, 6$\}$ & none  \\ 
$\{$4$\}$ & $\{$1, 3, 5$\}$ & partially exist &$\{$4$\}$ & $\{$1, 3, 6$\}$ & partially exist &$\{$4$\}$ & $\{$1, 5, 6$\}$ & partially exist  \\ 
$\{$4$\}$ & $\{$2, 3, 5$\}$ & partially exist &$\{$4$\}$ & $\{$2, 3, 6$\}$ & partially exist &$\{$4$\}$ & $\{$2, 5, 6$\}$ & partially exist  \\ 
$\{$4$\}$ & $\{$3, 5, 6$\}$ & always exist &$\{$5$\}$ & $\{$1, 2, 3$\}$ & none &$\{$5$\}$ & $\{$1, 2, 4$\}$ & partially exist  \\ 
$\{$5$\}$ & $\{$1, 2, 6$\}$ & partially exist &$\{$5$\}$ & $\{$1, 3, 4$\}$ & partially exist &$\{$5$\}$ & $\{$1, 3, 6$\}$ & partially exist  \\ 
$\{$5$\}$ & $\{$1, 4, 6$\}$ & partially exist &$\{$5$\}$ & $\{$2, 3, 4$\}$ & partially exist &$\{$5$\}$ & $\{$2, 3, 6$\}$ & partially exist  \\ 
$\{$5$\}$ & $\{$2, 4, 6$\}$ & partially exist &$\{$5$\}$ & $\{$3, 4, 6$\}$ & always exist &$\{$6$\}$ & $\{$1, 2, 3$\}$ & none  \\ 
$\{$6$\}$ & $\{$1, 2, 4$\}$ & none &$\{$6$\}$ & $\{$1, 2, 5$\}$ & partially exist &$\{$6$\}$ & $\{$1, 3, 4$\}$ & none  \\ 
$\{$6$\}$ & $\{$1, 3, 5$\}$ & partially exist &$\{$6$\}$ & $\{$1, 4, 5$\}$ & partially exist &$\{$6$\}$ & $\{$2, 3, 4$\}$ & none  \\ 
$\{$6$\}$ & $\{$2, 3, 5$\}$ & partially exist &$\{$6$\}$ & $\{$2, 4, 5$\}$ & partially exist &$\{$6$\}$ & $\{$3, 4, 5$\}$ & always exist  \\ 
\cline{1-9}
\end{tabular}
\end{table}

\begin{table}[h]
\begin{tabular}{||p{0.6cm}|p{1.6cm}|p{2cm}||p{0.6cm}|p{1.6cm}|p{2cm}||p{0.6cm}|p{1.6cm}|p{2cm}||}
\cline{1-9}
A               &   B           &   PPT Neg.     &   A          &    B           &   PPT Neg.        &  A           &    B           &   PPT Neg.  \\
\cline{1-9}
$\{$1$\}$ & $\{$2, 3, 4, 5$\}$ & always exist &$\{$1$\}$ & $\{$2, 3, 4, 6$\}$ & always exist &$\{$1$\}$ & $\{$2, 3, 5, 6$\}$ & always exist  \\ 
$\{$1$\}$ & $\{$2, 4, 5, 6$\}$ & always exist &$\{$1$\}$ & $\{$3, 4, 5, 6$\}$ & none &$\{$2$\}$ & $\{$1, 3, 4, 5$\}$ & always exist  \\ 
$\{$2$\}$ & $\{$1, 3, 4, 6$\}$ & always exist &$\{$2$\}$ & $\{$1, 3, 5, 6$\}$ & always exist &$\{$2$\}$ & $\{$1, 4, 5, 6$\}$ & always exist  \\ 
$\{$2$\}$ & $\{$3, 4, 5, 6$\}$ & partially exist &$\{$3$\}$ & $\{$1, 2, 4, 5$\}$ & always exist &$\{$3$\}$ & $\{$1, 2, 4, 6$\}$ & always exist  \\ 
$\{$3$\}$ & $\{$1, 2, 5, 6$\}$ & always exist &$\{$3$\}$ & $\{$1, 4, 5, 6$\}$ & always exist &$\{$3$\}$ & $\{$2, 4, 5, 6$\}$ & always exist  \\ 
$\{$4$\}$ & $\{$1, 2, 3, 5$\}$ & always exist &$\{$4$\}$ & $\{$1, 2, 3, 6$\}$ & always exist &$\{$4$\}$ & $\{$1, 2, 5, 6$\}$ & always exist  \\ 
$\{$4$\}$ & $\{$1, 3, 5, 6$\}$ & always exist &$\{$4$\}$ & $\{$2, 3, 5, 6$\}$ & always exist &$\{$5$\}$ & $\{$1, 2, 3, 4$\}$ & partially exist  \\ 
$\{$5$\}$ & $\{$1, 2, 3, 6$\}$ & always exist &$\{$5$\}$ & $\{$1, 2, 4, 6$\}$ & always exist &$\{$5$\}$ & $\{$1, 3, 4, 6$\}$ & always exist  \\ 
$\{$5$\}$ & $\{$2, 3, 4, 6$\}$ & always exist &$\{$6$\}$ & $\{$1, 2, 3, 4$\}$ & none &$\{$6$\}$ & $\{$1, 2, 3, 5$\}$ & always exist  \\ 
$\{$6$\}$ & $\{$1, 2, 4, 5$\}$ & always exist &$\{$6$\}$ & $\{$1, 3, 4, 5$\}$ & always exist &$\{$6$\}$ & $\{$2, 3, 4, 5$\}$ & always exist  \\ 
\cline{1-9}
\end{tabular}
\end{table}

\begin{table}[h]
\begin{tabular}{||p{1cm}|p{1.2cm}|p{2cm}||p{1cm}|p{1.2cm}|p{2cm}||p{1cm}|p{1.2cm}|p{2cm}||}
\cline{1-9}
A               &   B           &   PPT Neg.     &   A          &    B           &   PPT Neg.        &  A           &    B           &   PPT Neg.  \\
\cline{1-9}
$\{$1, 2$\}$ & $\{$3, 4$\}$ & always exist &$\{$1, 2$\}$ & $\{$3, 5$\}$ & partially exist &$\{$1, 2$\}$ & $\{$3, 6$\}$ & partially exist  \\ 
$\{$1, 2$\}$ & $\{$4, 5$\}$ & none &$\{$1, 2$\}$ & $\{$4, 6$\}$ & none &$\{$1, 2$\}$ & $\{$5, 6$\}$ & none  \\ 
$\{$1, 3$\}$ & $\{$2, 4$\}$ & always exist &$\{$1, 3$\}$ & $\{$2, 5$\}$ & partially exist &$\{$1, 3$\}$ & $\{$2, 6$\}$ & partially exist  \\ 
$\{$1, 3$\}$ & $\{$4, 5$\}$ & partially exist &$\{$1, 3$\}$ & $\{$4, 6$\}$ & partially exist &$\{$1, 3$\}$ & $\{$5, 6$\}$ & none  \\ 
$\{$1, 4$\}$ & $\{$2, 3$\}$ & always exist &$\{$1, 4$\}$ & $\{$2, 5$\}$ & partially exist &$\{$1, 4$\}$ & $\{$2, 6$\}$ & partially exist  \\ 
$\{$1, 4$\}$ & $\{$3, 5$\}$ & partially exist &$\{$1, 4$\}$ & $\{$3, 6$\}$ & partially exist &$\{$1, 4$\}$ & $\{$5, 6$\}$ & partially exist  \\ 
$\{$1, 5$\}$ & $\{$2, 3$\}$ & partially exist &$\{$1, 5$\}$ & $\{$2, 4$\}$ & partially exist &$\{$1, 5$\}$ & $\{$2, 6$\}$ & partially exist  \\ 
$\{$1, 5$\}$ & $\{$3, 4$\}$ & partially exist &$\{$1, 5$\}$ & $\{$3, 6$\}$ & partially exist &$\{$1, 5$\}$ & $\{$4, 6$\}$ & partially exist  \\ 
$\{$1, 6$\}$ & $\{$2, 3$\}$ & partially exist &$\{$1, 6$\}$ & $\{$2, 4$\}$ & none &$\{$1, 6$\}$ & $\{$2, 5$\}$ & partially exist  \\ 
$\{$1, 6$\}$ & $\{$3, 4$\}$ & none &$\{$1, 6$\}$ & $\{$3, 5$\}$ & partially exist &$\{$1, 6$\}$ & $\{$4, 5$\}$ & partially exist  \\ 
$\{$2, 3$\}$ & $\{$4, 5$\}$ & partially exist &$\{$2, 3$\}$ & $\{$4, 6$\}$ & partially exist &$\{$2, 3$\}$ & $\{$5, 6$\}$ & none  \\ 
$\{$2, 4$\}$ & $\{$3, 5$\}$ & partially exist &$\{$2, 4$\}$ & $\{$3, 6$\}$ & partially exist &$\{$2, 4$\}$ & $\{$5, 6$\}$ & partially exist  \\ 
$\{$2, 5$\}$ & $\{$3, 4$\}$ & partially exist &$\{$2, 5$\}$ & $\{$3, 6$\}$ & partially exist &$\{$2, 5$\}$ & $\{$4, 6$\}$ & partially exist  \\ 
$\{$2, 6$\}$ & $\{$3, 4$\}$ & partially exist &$\{$2, 6$\}$ & $\{$3, 5$\}$ & partially exist &$\{$2, 6$\}$ & $\{$4, 5$\}$ & partially exist  \\ 
$\{$3, 4$\}$ & $\{$5, 6$\}$ & always exist &$\{$3, 5$\}$ & $\{$4, 6$\}$ & always exist &$\{$4, 5$\}$ & $\{$3, 6$\}$ & always exist  \\ 
\cline{1-9}
\end{tabular}
\end{table}

\begin{table}[h]
\begin{tabular}{||p{1cm}|p{1.2cm}|p{2cm}||p{1cm}|p{1.2cm}|p{2cm}||p{1cm}|p{1.2cm}|p{2cm}||}
\cline{1-9}
A               &   B           &   PPT Neg.     &   A          &    B           &   PPT Neg.        &  A           &    B           &   PPT Neg.  \\
\cline{1-9}
$\{$1, 2$\}$ & $\{$3, 4, 5$\}$ & always exist &$\{$1, 2$\}$ & $\{$3, 4, 6$\}$ & always exist &$\{$1, 2$\}$ & $\{$3, 5, 6$\}$ & always exist  \\ 
$\{$1, 2$\}$ & $\{$4, 5, 6$\}$ & always exist &$\{$1, 3$\}$ & $\{$2, 4, 5$\}$ & always exist &$\{$1, 3$\}$ & $\{$2, 4, 6$\}$ & always exist  \\ 
$\{$1, 3$\}$ & $\{$2, 5, 6$\}$ & always exist &$\{$1, 3$\}$ & $\{$4, 5, 6$\}$ & always exist &$\{$1, 4$\}$ & $\{$2, 3, 5$\}$ & always exist  \\ 
$\{$1, 4$\}$ & $\{$2, 3, 6$\}$ & always exist &$\{$1, 4$\}$ & $\{$2, 5, 6$\}$ & always exist &$\{$1, 4$\}$ & $\{$3, 5, 6$\}$ & always exist  \\ 
$\{$1, 5$\}$ & $\{$2, 3, 4$\}$ & always exist &$\{$1, 5$\}$ & $\{$2, 3, 6$\}$ & always exist &$\{$1, 5$\}$ & $\{$2, 4, 6$\}$ & always exist  \\ 
$\{$1, 5$\}$ & $\{$3, 4, 6$\}$ & always exist &$\{$1, 6$\}$ & $\{$2, 3, 4$\}$ & always exist &$\{$1, 6$\}$ & $\{$2, 3, 5$\}$ & always exist  \\ 
$\{$1, 6$\}$ & $\{$2, 4, 5$\}$ & always exist &$\{$1, 6$\}$ & $\{$3, 4, 5$\}$ & always exist &$\{$2, 3$\}$ & $\{$1, 4, 5$\}$ & always exist  \\ 
$\{$2, 3$\}$ & $\{$1, 4, 6$\}$ & always exist &$\{$2, 3$\}$ & $\{$1, 5, 6$\}$ & always exist &$\{$2, 3$\}$ & $\{$4, 5, 6$\}$ & always exist  \\ 
$\{$2, 4$\}$ & $\{$1, 3, 5$\}$ & always exist &$\{$2, 4$\}$ & $\{$1, 3, 6$\}$ & always exist &$\{$2, 4$\}$ & $\{$1, 5, 6$\}$ & always exist  \\ 
$\{$2, 4$\}$ & $\{$3, 5, 6$\}$ & always exist &$\{$2, 5$\}$ & $\{$1, 3, 4$\}$ & always exist &$\{$2, 5$\}$ & $\{$1, 3, 6$\}$ & always exist  \\ 
$\{$2, 5$\}$ & $\{$1, 4, 6$\}$ & always exist &$\{$2, 5$\}$ & $\{$3, 4, 6$\}$ & always exist &$\{$2, 6$\}$ & $\{$1, 3, 4$\}$ & always exist  \\ 
$\{$2, 6$\}$ & $\{$1, 3, 5$\}$ & always exist &$\{$2, 6$\}$ & $\{$1, 4, 5$\}$ & always exist &$\{$2, 6$\}$ & $\{$3, 4, 5$\}$ & always exist  \\ 
$\{$3, 4$\}$ & $\{$1, 2, 5$\}$ & always exist &$\{$3, 4$\}$ & $\{$1, 2, 6$\}$ & always exist &$\{$3, 4$\}$ & $\{$1, 5, 6$\}$ & always exist  \\ 
$\{$3, 4$\}$ & $\{$2, 5, 6$\}$ & always exist &$\{$3, 5$\}$ & $\{$1, 2, 4$\}$ & always exist &$\{$3, 5$\}$ & $\{$1, 2, 6$\}$ & always exist  \\ 
$\{$3, 5$\}$ & $\{$1, 4, 6$\}$ & always exist &$\{$3, 5$\}$ & $\{$2, 4, 6$\}$ & always exist &$\{$3, 6$\}$ & $\{$1, 2, 4$\}$ & always exist  \\ 
$\{$3, 6$\}$ & $\{$1, 2, 5$\}$ & always exist &$\{$3, 6$\}$ & $\{$1, 4, 5$\}$ & always exist &$\{$3, 6$\}$ & $\{$2, 4, 5$\}$ & always exist  \\ 
$\{$4, 5$\}$ & $\{$1, 2, 3$\}$ & always exist &$\{$4, 5$\}$ & $\{$1, 2, 6$\}$ & always exist &$\{$4, 5$\}$ & $\{$1, 3, 6$\}$ & always exist  \\ 
$\{$4, 5$\}$ & $\{$2, 3, 6$\}$ & always exist &$\{$4, 6$\}$ & $\{$1, 2, 3$\}$ & always exist &$\{$4, 6$\}$ & $\{$1, 2, 5$\}$ & always exist  \\ 
$\{$4, 6$\}$ & $\{$1, 3, 5$\}$ & always exist &$\{$4, 6$\}$ & $\{$2, 3, 5$\}$ & always exist &$\{$5, 6$\}$ & $\{$1, 2 ,3$\}$ & always exist  \\ 
$\{$5, 6$\}$ & $\{$1, 2, 4$\}$ & always exist &$\{$5, 6$\}$ & $\{$1, 3, 4$\}$ & always exist &$\{$5, 6$\}$ & $\{$2, 3, 4$\}$ & always exist  \\   
\cline{1-9}
\end{tabular}
\end{table}